\documentclass[bibyear]{aa}
\usepackage[varg]{txfonts}
\usepackage{graphicx}
\def\I{\mbox{IRAS\,20126+4104}}

\def\Mc{M_{\rm c}}

\def\Msun{\mbox{$M_\odot$}}
\def\Lsun{\mbox{$L_\odot$}}
\def\WAT{H$_2$O}
\def\CO{$^{12}$CO}
\def\COI{\mbox{$^{13}$CO}}

\def\OI{\mbox{O{\sc i}}}
\def\CII{\mbox{C{\sc ii}}}

\def\HM{\mbox{H$_2$}}

\def\kms{\mbox{km~s$^{-1}$}}
\def\cmc{cm$^{-3}$}
\def\cmq{cm$^{-2}$}

\def\mic{\mbox{$\mu$m}}
\def\Log{\mbox{\rm log$_{10}$}}

\def\d{{\rm d}}
\def\e{{\rm e}}

\def\nHM{\mbox{$n_{\rm H_2}$}}
\def\No{\mbox{$N_{\rm O}$}}
\def\NCO{N_{\rm CO}}

\def\Omc{\Omega_{\rm c}}

\def\To{T_{\rm o}}
\def\no{n_{\rm o}}
\def\Ro{R_{\rm o}}
\def\Ri{R_{\rm i}}

\begin{document}

\title{
A Herschel study of the high-mass protostar \I
}
\author{
        R.~Cesaroni\inst{1} \and\ F.~Faustini\inst{2,3} \and\ D.~Galli\inst{1} \and\ A.~Lorenzani\inst{1} \and\ S.~Molinari\inst{4} \and\ L.~Testi\inst{5,1,6}
}
\institute{
 INAF, Osservatorio Astrofisico di Arcetri, Largo E. Fermi 5, I-50125 Firenze, Italy
           \email{riccardo.cesaroni@inaf.it}
 \and
 ASI -- Space Science Data Center, via del Politecnico, 00133, Roma, Italy
 \and
 INAF -- Osservatorio Astronomico di Roma, Via Frascati 33, 00078 Monteporzio Catone, Rome, Italy
 \and
 INAF -- Istituto di Astrofisica e Planetologia Spaziali, Via Fosso del Cavaliere 100, I-00133, Rome, Italy
 \and
 European Southern Observatory, Karl-Schwarzschild-Str. 2, 85748
 Garching bei M\"unchen, Germany
 \and
 Alma Mater Studiorum Universit\'a di Bologna, Dipartimento di Fisica e Astronomia (DIFA), Via Gobetti 93/2, I-40129 Bologna, Italy
}
\offprints{R. Cesaroni, \email{riccardo.cesaroni@inaf.it}}
\date{Received date / Accepted date}

\abstract{
We performed Herschel observations of the continuum and line emission from
the high-mass star-forming region \I, which hosts a well-studied B-type
(proto)star powering a bipolar outflow and is associated with a Keplerian
circumstellar disk. The continuum images at six wavelengths allowed
us to
derive an accurate estimate of the bolometric luminosity and mass of
the molecular clump enshrouding the disk.
The same region has been mapped in 12 rotational transitions of carbon
monoxide, which were used in synergy with the continuum data to
determine the temperature and density distribution inside the clump and
improve upon the mass estimate.
The maps of two
fine structure
oxygen far-IR
lines were used to estimate the volume density of the shocked region at the
surface of the southern lobe of the outflow and
the mass-loss rate. Our findings lend further support to the scenario
previously proposed by various authors, confirming that at the origin of
the bolometric luminosity and bipolar outflow from \I\ is a B-type
star located at the centre of the Keplerian disk.
}

\maketitle

\section{Introduction}
\label{sint}

The study of massive stars (conventionally those in excess of $\sim$8~\Msun)
is hampered by various problems, the most important of which probably
being their large distances and their formation in clusters. As a consequence, precise
estimates of crucial physical quantities such as the stellar luminosity in addition to
accretion and outflow rates had been difficult to obtain until recently. With the
substantial improvement of sensitivity and angular resolution provided by
the new instruments that have recently become available online, these limitations can be overcome, at least
in part. In this study we exploit the potential of
the ESA Herschel Space Observatory (Pilbratt et al. \cite{pilb})
to image the continuum and line emission from a massive (proto)star
in the far-IR at a few 10\arcsec\ resolution
in the wake of previous studies by other authors (e.g. Green et
al.~\cite{green13}; Manoj et al.~\cite{manoj13,manoj16}; Karska et
al.~\cite{karska13,karska14,karska18}; Nisini et al.~\cite{nisi}; Mottram
et al.~\cite{mott17}; Yang et al.~\cite{yang18}). In particular, the fine
structure \OI\ line at 63~\mic\ has been found to be tightly associated
with outflows in both low-mass (Kristensen et al.~\cite{krist17}; Yang et
al.~\cite{yang22}) and high-mass (Schneider et al.~\cite{schnei18}) young
stellar objects (YSOs), although it is possibly affected (especially at low
velocities) by absorption in foreground clouds (Leurini et al.~\cite{leu15}).

The target of our study is the well-known massive (proto)star \I. This object
was probably the first Keplerian disk found around a massive (proto)star and
it has been extensively studied over almost the entire electromagnetic spectrum
accessible to the observers, from the radio to the X-ray domain. While it
would be too lengthy to review all the results obtained, here we mention the
most relevant to the present study.

The (proto)star lies at the centre of a disk,  which was first
 detected by Cesaroni et al.~(\cite{cesa97}) and confirmed by subsequent higher-resolution studies (Zhang et al.~\cite{zha98}; Cesaroni et al.~\cite{cesa99,cesa05,cesa14}).
Sub-arcsecond imaging (Cesaroni et al.~\cite{cesa14}) and model fitting
 (Chen et al.~\cite{chen}) have proved that the disk is undergoing
 Keplerian rotation about a $\sim$$12~M_\odot$ (proto)star, as suggested by the
 butterfly-shaped position--velocity plot obtained in almost any hot molecular
 core tracer observed. Evidence of
 accretion onto the (proto)star has also been reported (Cesaroni et al.~\cite{cesa99};
 Keto \& Zhang~\cite{ketzha}; Johnston et al.~\cite{johns}).
The (proto)star is powering a outflow undergoing
 precession about the disk axis (Shepherd et al.~\cite{shep}; Cesaroni et al.~\cite{cesa05};
 Caratti o Garatti et al.~\cite{caga08}).
 This outflow has been imaged on scales from a few
 100~au (Moscadelli et al.~\cite{mosca05}; Cesaroni et al.~\cite{cesa13}) to 0.5~pc (Shepherd et
 al.~\cite{shep}; Lebr\'on et al.~\cite{lebr}) and its 3D expansion velocity has been
 measured through maser and \HM-knot proper motions (Moscadelli et al.~\cite{mosca05}; Massi et al.~\cite{massi23}).
The distance to \I\ (1.64$\pm$0.05~kpc) has been accurately
 determined from parallax measurements of the \WAT\ masers (Moscadelli et
 al.~\cite{mosca11}), which prove this to be one of the closest disks around B-type
 (proto)stars, thus allowing for excellent linear resolution. More recently,
 Nagayama et al.~(\cite{naga15}) have repeated the parallax measurement with
 the Japanese Very-Long-Baseline Interferometry (VLBI) Exploration of Radio Astrometry (VERA)
 resulting in a distance of 1.33$^{+0.19}_{-0.092}$~kpc, consistent with the
 value of Moscadelli et al. within the uncertainties. For our study we adopt
 the distance estimate with the minimum error, that is 1.64~kpc.
Imaging at IR wavelengths (Qiu et al.~\cite{qiu08}) seemed to indicate
 that \I\ is associated with an anomalously poor stellar cluster. However,
 subsequent X-ray observations by Montes et al.~(\cite{montes}) have revealed an
 embedded stellar population that hints at the existence of a richer
 (and possibly very young) cluster.

The aim of this paper is to obtain precise estimates of the luminosity and
outflow mass-loss rate, which are necessary to
shed light on the stellar properties and possible stellar multiplicity.
We also want to establish the origin of the \OI\ line emission and
use it to obtain an alternative estimate of the mass-loss
rate in the jet.
With this in mind, we performed Herschel observations of the continuum
and line emission from \I.

This paper is organized as follows. In Sect.~\ref{sobs} the observations
and data reduction procedures are described. In Sect.~\ref{sres} we outline
the results obtained, which are then analysed and discussed in Sect.~\ref{sdis}.
Finally, the conclusions are drawn in Sect.~\ref{ssum}.

\section{Observations and data reduction}
\label{sobs}

We observed a total time of 11.5 hours with the
Photodetector Array Camera and Spectrometer (PACS; Poglitsch et
al.~\cite{pogl}) and the Spectral and Photometric Imaging Receiver (SPIRE; Griffin et al.~\cite{griff}),
to cover a region of at least 2\arcmin$\times$2\arcmin\ centred on \I.
The photometric and spectroscopic data are described separately in the
following sections.

\subsection{Photometry}

\subsubsection{Data acquisition}

We obtained observations in photometric mode with both PACS and SPIRE of
an area similar to that mapped in spectroscopic mode. Continuum maps of
the region were made in six bands at 70, 100, 110, 160, 250, 350,
and 500~\mic. A single coverage was sufficient at all bands due to the
strong source intensity.

SPIRE was used in mini-map mode, employing a small scan map using the nearly orthogonal
scan paths, with a fixed scan angle (42\degr\ with respect to the z-axis of the
spacecraft) and scan speed (30\arcsec/s), resulting in a circular coverage of
5\arcmin\ diameter.
PACS was used in scan-map mode to map a region of about
3\arcmin$\times$3\arcmin. Two coverages were necessary for all three PACS
bands, implying a time of 558~sec. A mini map was obtained by moving the
array with a constant speed of 20\arcsec/s along four
parallel legs, with small
leg separation (2\arcsec--5\arcsec). The length of the legs was 3\arcmin.
Scanning was performed in array coordinates at 70\degr/110\degr, in order to
align the scan direction along the diagonal of the array. To improve the
sensitivity and better image extended sources, two consecutive orthogonal mini
scan-maps were acquired. The obtained map shows a central homogeneous,
high-coverage area with a diameter of 50\arcsec. The execution time of a
mini map composed by the concatenation of two quasi-orthogonal directions was
567 second (for six scan legs with a separation of 4\arcsec\ and leg length
of 3\arcmin), of which 104 seconds were effectively spent on the source.

\subsubsection{Data reduction}
\label{photdataq}

The standard products were obtained by performing queries at the Herschel
Science Archive (HSA), using jython (java+python) scripts developed within
the Herschel Interactive Processing Environment (HIPE\footnote{HIPE is a
joint development by the Herschel Science Ground
Segment Consortium, consisting of ESA, the NASA Herschel Science Center,
and the HIFI, PACS, and SPIRE consortia.}; Ott~\cite{ott}). The level 2
standard products were extracted from the observational context and were
processed with dedicated scripts to optimize the results and completely
remove the residual features.

The raw data acquired with the SPIRE mini-map mode,
at 250, 350, and 500~\mic,
contain both orthogonal
scan directions in a single observation. The standard pipeline combines
all the scan directions, producing a single observation. For our study,
we used the level 2 standard products.

For maps acquired with PACS, every scan direction corresponds to a single
observation; therefore, all of the observations had to be combined to obtain
a map. For this purpose, a specific pipeline was created that uses the level 0 time ordered data (TOD) from the HSA, after removing all of the
instrumental and observational effects, such as Glitch, drift, offset, and
1/f noise. Our pipeline produces one map for each PACS band (70, 100, 110,
and 160 $\mu$m). For our study we used these reprocessed data.

\subsection{Spectroscopy}

\subsubsection{Data acquisition}

PACS was used in spectroscopic mode to build a raster map of a region
of 2\arcmin$\times$2\arcmin\ centred on RA=20$^{\rm h}$14$^{\rm
m}$25\fs5 Dec=41\degr13\arcmin10\arcsec\ in 
five
lines: \OI\ 63~$\mu$m, \OI\ 145~$\mu$m, \CII\ 157~$\mu$m,
\CO\ $J$=18$\rightarrow$17 at 145~$\mu$m,
H$_2$O $2_{2,1}$--$1_{1,0}$ at 108~\mic,
and \CO\ $J$=24$\rightarrow23$ at 109~$\mu$m. The total integration time
was 14867~sec.
SPIRE was used in single pointing spectral mode with an intermediate
coverage and high resolution to obtain the interferogram in the band from
194~$\mu$m to 672~$\mu$m, with the aim to map the same area as PACS in all
\CO\ rotational transitions between $J$=13$\rightarrow$12, at 200~$\mu$m,
and $J$=4$\rightarrow$3, at 650~$\mu$m.

\subsubsection{Data reduction}

With PACS, the lines were observed through three distinct
observations using the observing mode \texttt{Mapping,Chop/Nod}. 
In this mode, for each ON position, there are two OFF positions.
For some of the observations, the standard product obtained from
the Herschel Science Archive (see Sect.~\ref{photdataq})
presents absorption lines in regions where no emission is detected.
These are due to the presence of emission lines in one or both OFF position(s). 
Thus we used specific scripts to remove such lines 
together with the official PACS pipeline to reduce the observations.
The latest HIPE version 16.365 was used to reprocess the data.

Observation n.~1342234936 covers the \CO\ $J$=24$\rightarrow23$
and H$_2$O $2_{2,1}$--$1_{1,0}$ lines in the red band of the spectrometer,
while no line was detected in the blue band.
For
this observation no absorption feature was detected and in fact a check of
the spectra in the OFF positions confirms the absence of emission lines.
The standard pipeline was used to reduce the data and obtain the spectral
cube.

Observation n.~1342235689 covers the \OI\ 145~$\mu$m and
\CO(18--17)
lines in the red band of the spectrometer. This
observation presents an absorption line in the southern part of
the mapped region and an analysis of the OFF positions through the
\texttt{ChopNodSplitOnOff} script provided in the \texttt{HIPE -- PACS
pipeline Script} section confirms the presence of an emission line in
the OFF B position (maximum at RA=303.48038~deg, Dec=41.25495~deg with
intensity $\sim$0.45~Jy). To remove this line, we used a slightly modified
version of the PACS pipeline. In particular, we ran the pipeline script that
reprocesses the data from level 0 up to the re-binned cubes, then we retrieved
the spectra related to the OFF B position and removed the emission with
a linear interpolation of the continuum 
on the two sides of
the line. Then
the modified spectra were re-inserted in the original cube and  processed
with the rest of the pipeline.

Observation n.~1342235686 covers the \CII\
line
in the red band of the
spectrometer, and the \OI\ 63~$\mu$m line in the blue band. Although no
obvious absorption feature was seen in the mapped region, a detailed check
of the spectra unveiled diffuse line emission at the \CII\ wavelength in
both OFF positions with intensities varying from 0.8 to 1.4~Jy
and from 2 to 8~Jy, respectively.  A procedure similar to the one used for observation
n.~1342235689 was applied to remove the emission line in the OFF positions
and apply the corresponding correction to the data.  The final difference
between the line extracted from the original product and the line obtained
with the procedure described above was at most $\sim$1Jy.

All data were then transformed into GILDAS\footnote{The GILDAS
software has been developed at IRAM and the Observatoire de Grenoble:
http://www.iram.fr/IRAMFR/GILDAS}
format through the following steps.
Using HIPE,
the cubes were treated in such a way to convert the wavelength
axes into velocities at the rest frequency of the lines.
Then the resulting cubes were resampled so that the channel width was constant.
The continuum was removed with the task \texttt{removeBaselineFromCube}.
Then the cubes were cropped around the lines and exported in fits files 
with the task \texttt{simpleFitsWriter}. Finally, these files were
converted into GILDAS format with the standard GILDAS procedures.

With SPIRE, all lines were covered in a single observation n.~1342243604,
using a \texttt{Single Pointing -- Intermediate Map Sampling} mode, with
\texttt{HR} spectral resolution in detector bands SLW and SSW.  In each
band the apodized cube continumm obtained from the SPIRE pipeline was
subtracted with a modified version of the \texttt{Spire Useful Script
-- Spectrometer Cube Fitting}. Then, for each line, the cube wavelength
axis was converted into a velocity axis and cropped around the line.
The cropped cubes as well as the extracted central spectra were exported
in FITS format and then read into GILDAS as for the PACS data.

\begin{figure*}
\centering
\resizebox{16.5cm}{!}{\includegraphics[angle=0]{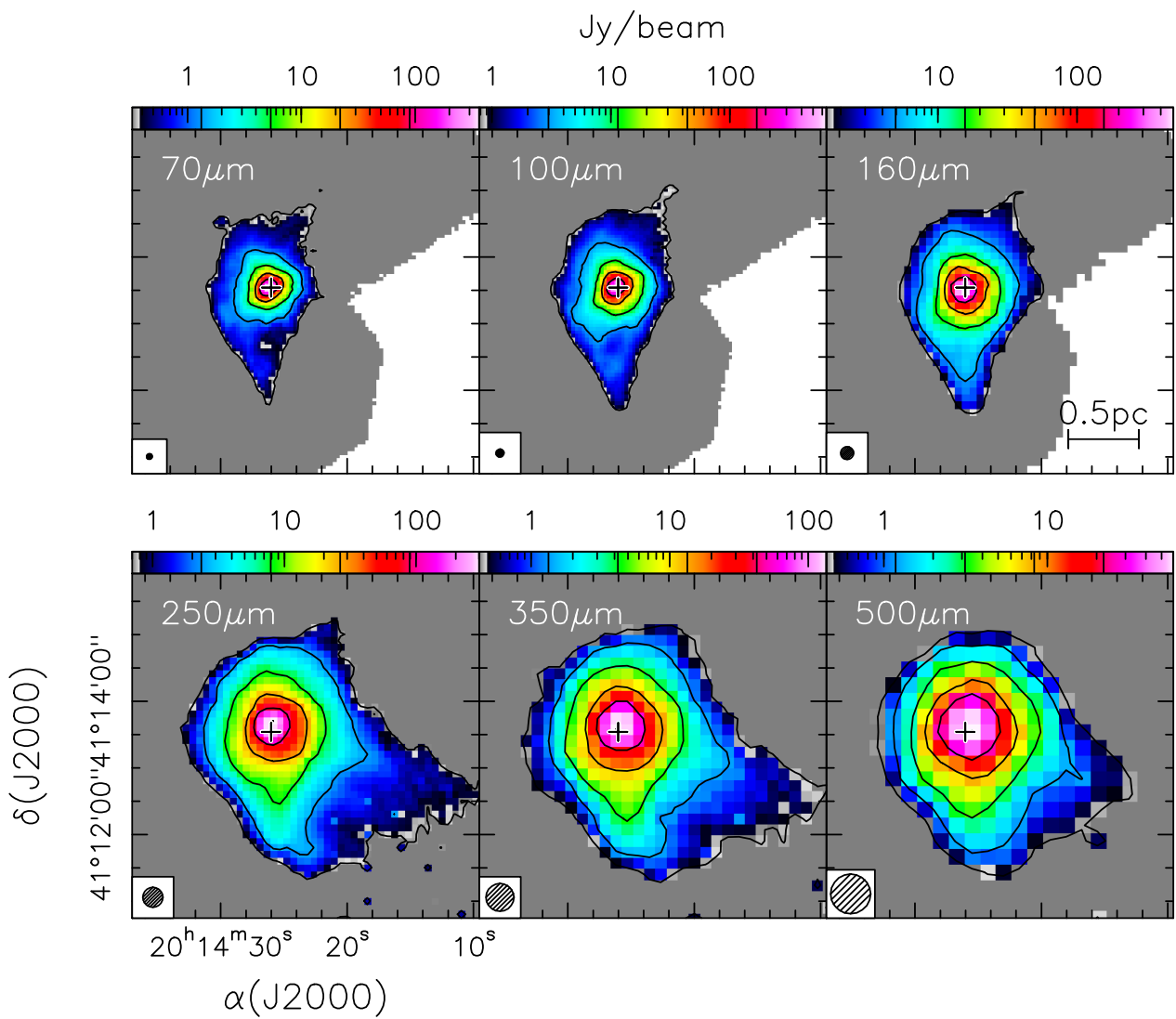}}
\caption{
Maps of the \I\ region obtained with Herschel at different wavelengths,
as indicated in each panel.
The cross marks the position of the circumstellar disk.
The circles in the bottom left denote the
full-width at half power of the instrumental beam. The values of the contour
levels are marked by vertical bars in the corresponding colour scale.
}
\label{fmcon}
\end{figure*}

\begin{figure*}
\centering
\resizebox{16.5cm}{!}{\includegraphics[angle=0]{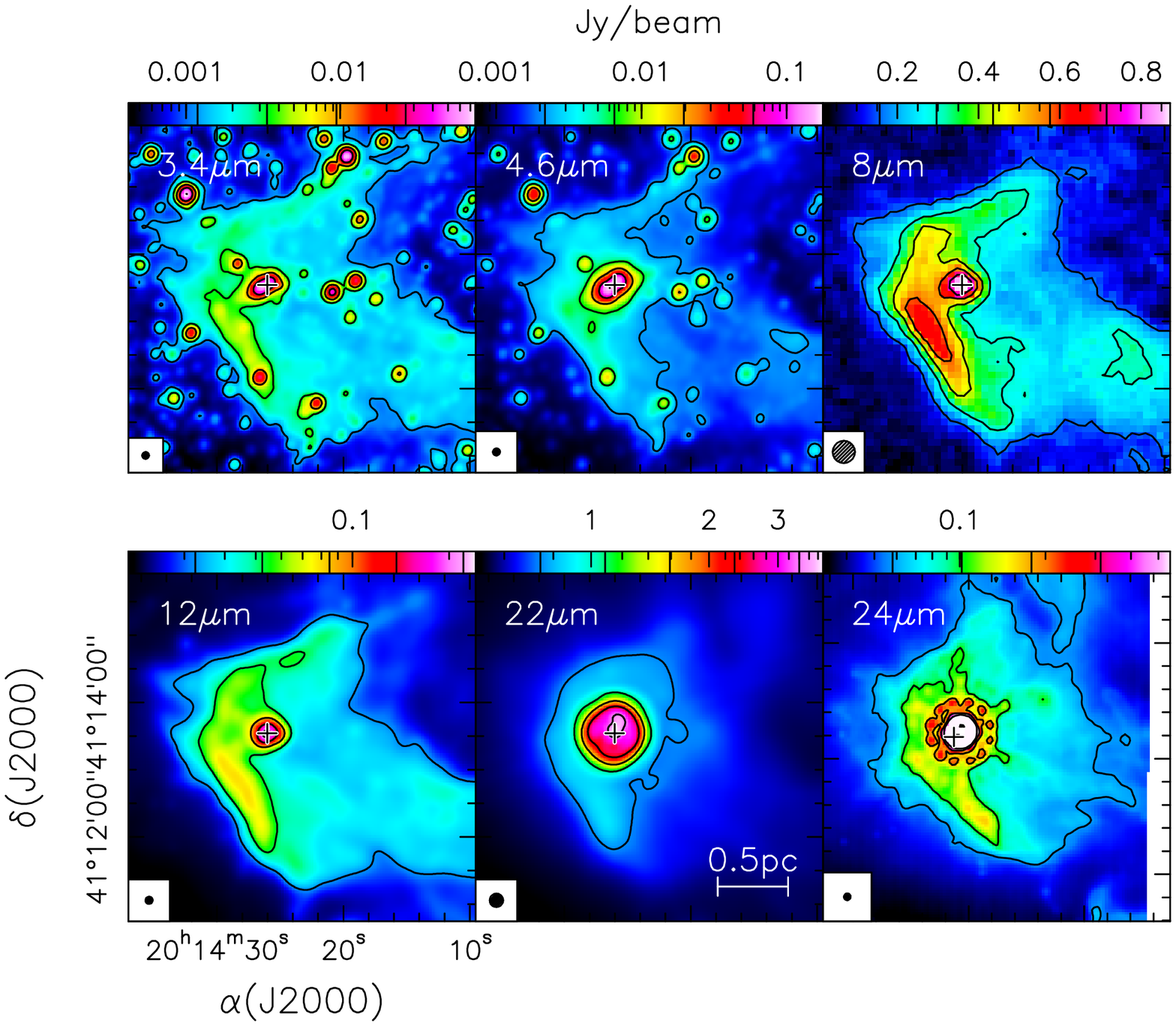}}
\caption{
Same as Fig.~\ref{fmcon}, but for archival images in the IR. The images at 3.4,
4.6, 12, and 22~$\mu$m are from the WISE survey, while those at 8~$\mu$m
and 24~$\mu$m are from the MSX survey and the Spitzer database, respectively.
The latter image is heavily saturated and shows the pattern of the point
spread function of the telescope.
}
\label{fmir}
\end{figure*}

\section{Results}
\label{sres}

\subsection{Continuum}
\label{scon}

The maps of the continuum emission obtained at the six observed bands
(70, 100, 160, 250, 350, and 500~\mic)
are shown in Fig.~\ref{fmcon}. In all cases, a clear intensity peak is
visible at the position of the circumstellar disk, marked by a cross
in the figure. While at the shortest wavelengths, the emission is quite
compact; in the sub-millimetre regime, one can see an extension to the south-west,
likely associated with colder dust.

For the sake of completeness, we have also retrieved archival IR images
at shorter wavelengths of the same region mapped with Herschel. These
are shown in Fig.~\ref{fmir}. More precisely, we have used data from
the Midcourse Space Experiment (MSX) (Price et al.~\cite{price}), the
Wide-field Infrared Survey Explorer (WISE) (Wright et al.~\cite{wright}),
and the Spitzer Space Telescope database
(Werner et al.~\cite{werner}; Fazio et al.~\cite{fazio}; Rieke et al.~\cite{rieke})
Unlike the maps of Fig.~\ref{fmcon},
the structure of the source appears less concentrated and, in
particular, an elongated region of emission is seen to the south-east,
probably due to a photo-dissociation region (PDR), as indicated especially
by the MSX image at 8~\mic, a wavelength at which PAH emission is very prominent
(see, e.g. Tielens~\cite{tiel}).

All in all, the continuum emission from \I\ is relatively simple and appears
to be dominated by one or more sources concentrated in the central region of a
parsec-scale molecular, dusty clump. We can thus derive a reliable estimate
of the luminosity of \I, as done in the following (see Sect.~\ref{slum}).

\subsection{Line}
\label{slin}

\begin{figure*}
\centering
\resizebox{16.0cm}{!}{\includegraphics[angle=0]{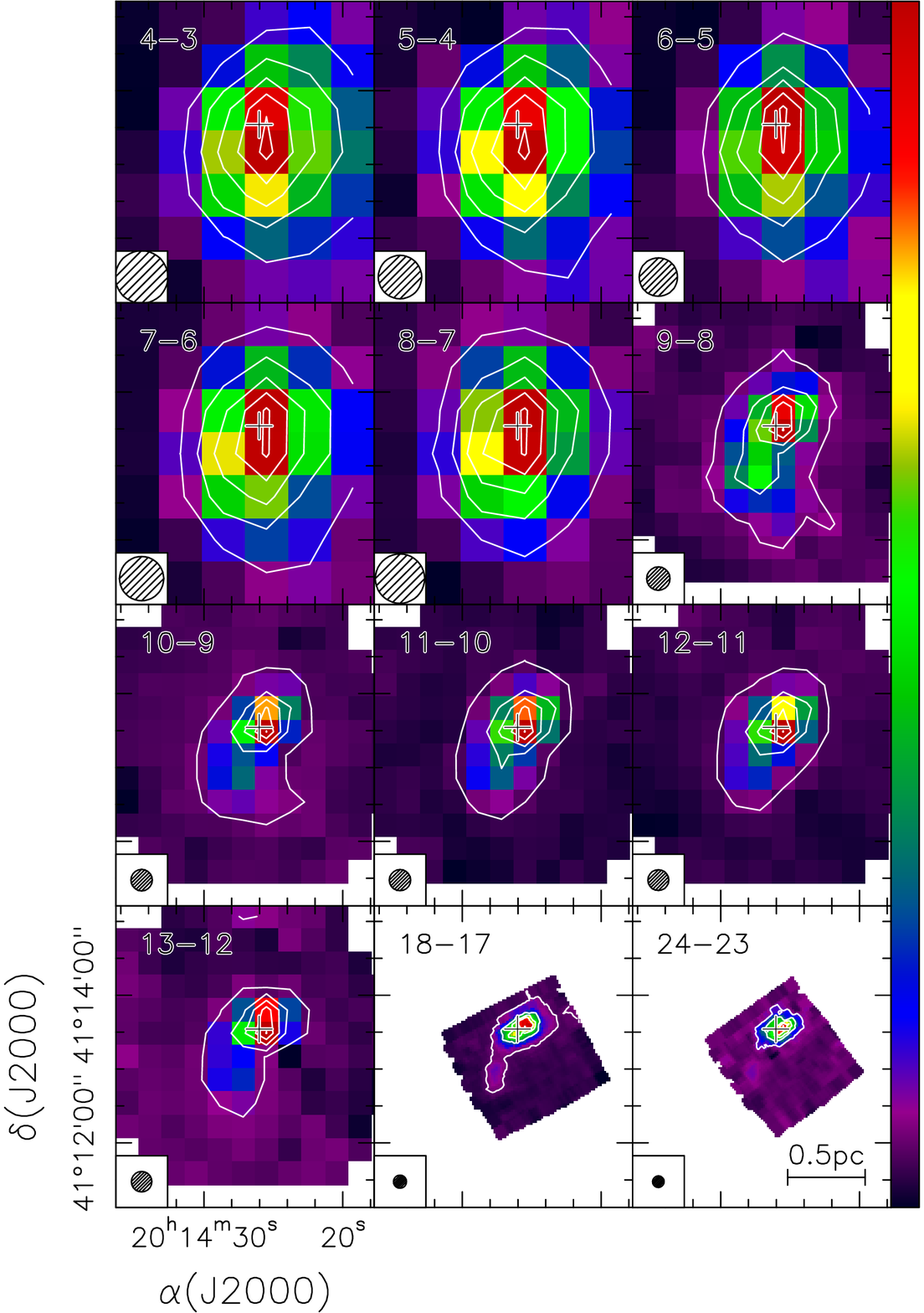}}
\caption{
Maps of the emission averaged over the observed \CO\ lines. The numbers
in the top left of each panel indicate the $J+1\rightarrow J$ rotational
transition, while the circle in the bottom left denotes the corresponding
HPBW. The cross marks the position of the disk. The minimum, maximum, and step
for contour levels in units of Jy/beam are as follows (panel by panel,
from top to bottom, and from left to right):
3, 15.99, 3.25;
3, 19.63, 4.16;
3, 19.58, 4.14;
3, 25.79, 5.7;
3, 26.51, 5.88;
0.6, 15.98, 3.85;
0.6, 22.89, 5.57;
0.6, 21.25, 5.16;
0.6, 21.42, 5.2;
0.6, 16.99, 4.1;
0.39, 14.13, 3.43;
0.24, 3.69, 0.86.
}
\label{fmco}
\end{figure*}

\begin{figure*}
\centering
\resizebox{16.5cm}{!}{\includegraphics[angle=0]{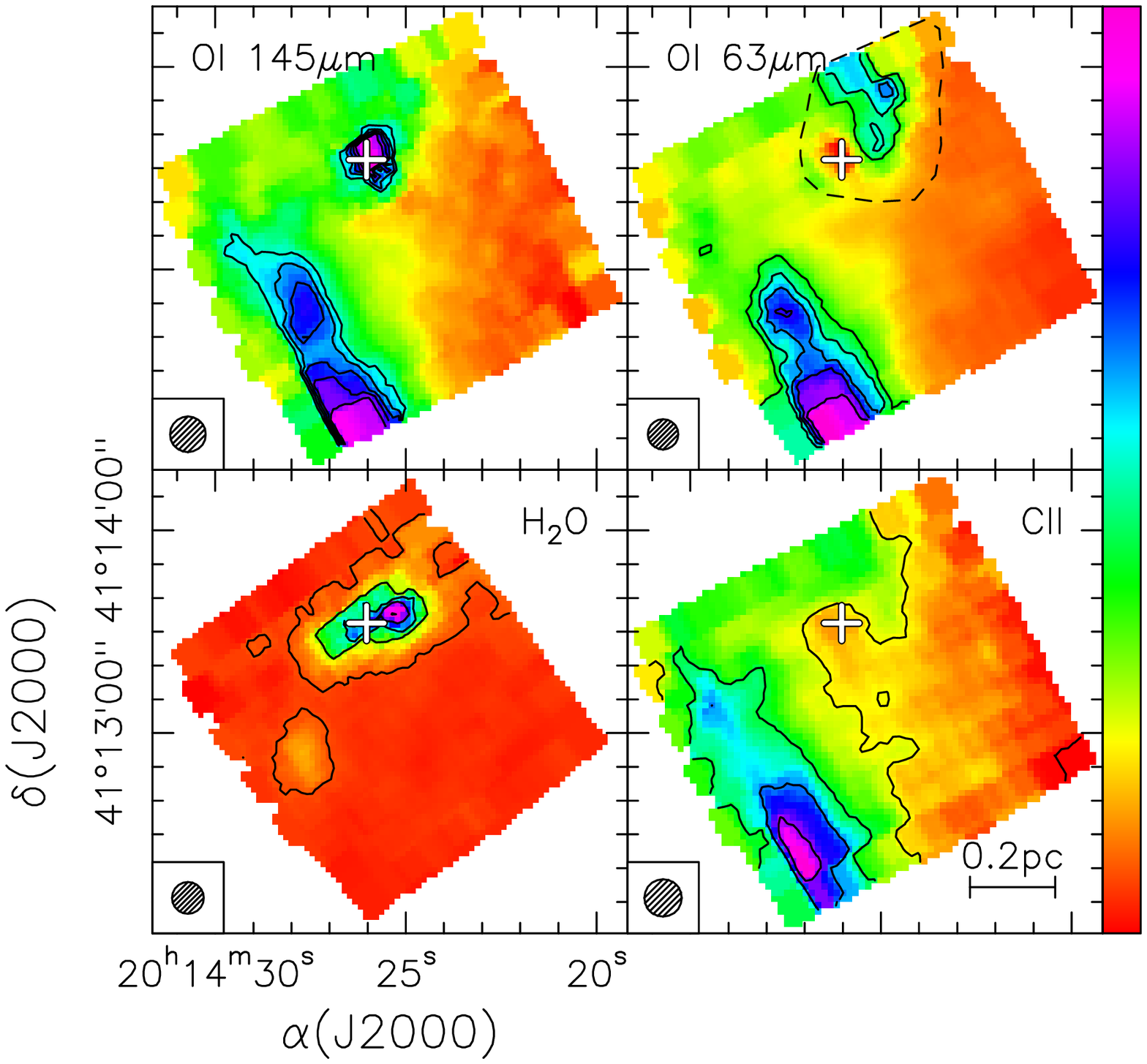}}
\caption{
Same as Fig.~\ref{fmco}, but for the \WAT, \OI, and \CII\ lines.
The dashed polygon outlines the region where an artefact is present
in the \OI\ 63~\mic\ line emission.
The minimum, maximum, and step for contour levels in units of Jy/beam are as
follows (panel by panel, from top to bottom, and from left to right):
1.44, 2.27, 0.21;
13.87, 29.01, 3.78;
0.32, 9.27, 2.24;
32.27, 87.41, 13.78.
}
\label{fmoch}
\end{figure*}

As detailed in Sect.~\ref{sobs}, the chosen spectral setup allowed us to
cover all carbon monoxide rotational transitions from (4--3) to (13--12)
as well as the
\CO (18--17)
and (24--23) lines. At the same time, we
observed the oxygen $^3P_1$--$^3P_2$ and $^3P_0$--$^3P_1$ transitions at
63.18~\mic\ and 145.53~\mic, 
respectively, and the $[$\CII$]$
$^2P_{3/2}$--$^2P_{1/2}$ transition at 157.74~\mic. We
also detected the water $2_{2,1}$--$1_{1,0}$ line at 108.07~\mic,
which happens
to fall in the same band as the \CO(24--23) transition. Unfortunately, in
all cases the spectral resolution is largely insufficient to resolve the line
profiles and obtain information on the kinematics of the gas. Therefore,
in our study we consider only the emission averaged over the lines, whose
maps are shown in Figs.~\ref{fmco} and~\ref{fmoch}.

The most striking difference between the molecular (\CO\ and \WAT) and atomic
(\OI\ and \CII) lines is that the former are basically concentrated around
the position of the disk, whereas the
latter appear to trace the south-eastern edge of the surrounding parsec-scale
clump.
We note that the spectral resolution of our observations is not
sufficient to resolve the CO line wings and thus it cannot image the outflow lobes
previously identified by Shepherd et al.~(\cite{shep}) and Lebr\'on et
al.~(\cite{lebr}). Our maps basically trace the bulk emission close to
the systemic velocity, which is likely to dominate over the wing emission
even in the high energy transitions -- as suggested by the \CO(7--6)
spectrum of Kawamura et al.~(\cite{kawa99}). The same conclusion might
hold for the \WAT\ emission: although no bipolar structure in seen
in the corresponding map of Fig.~\ref{fmoch}, one cannot rule out the
possibility that part of the emission arises from the outflow, as found
in low-mass protostars (van Dishoeck et al.~\cite{vandish}).

\begin{figure*}
\centering
\resizebox{16.5cm}{!}{\includegraphics[angle=0]{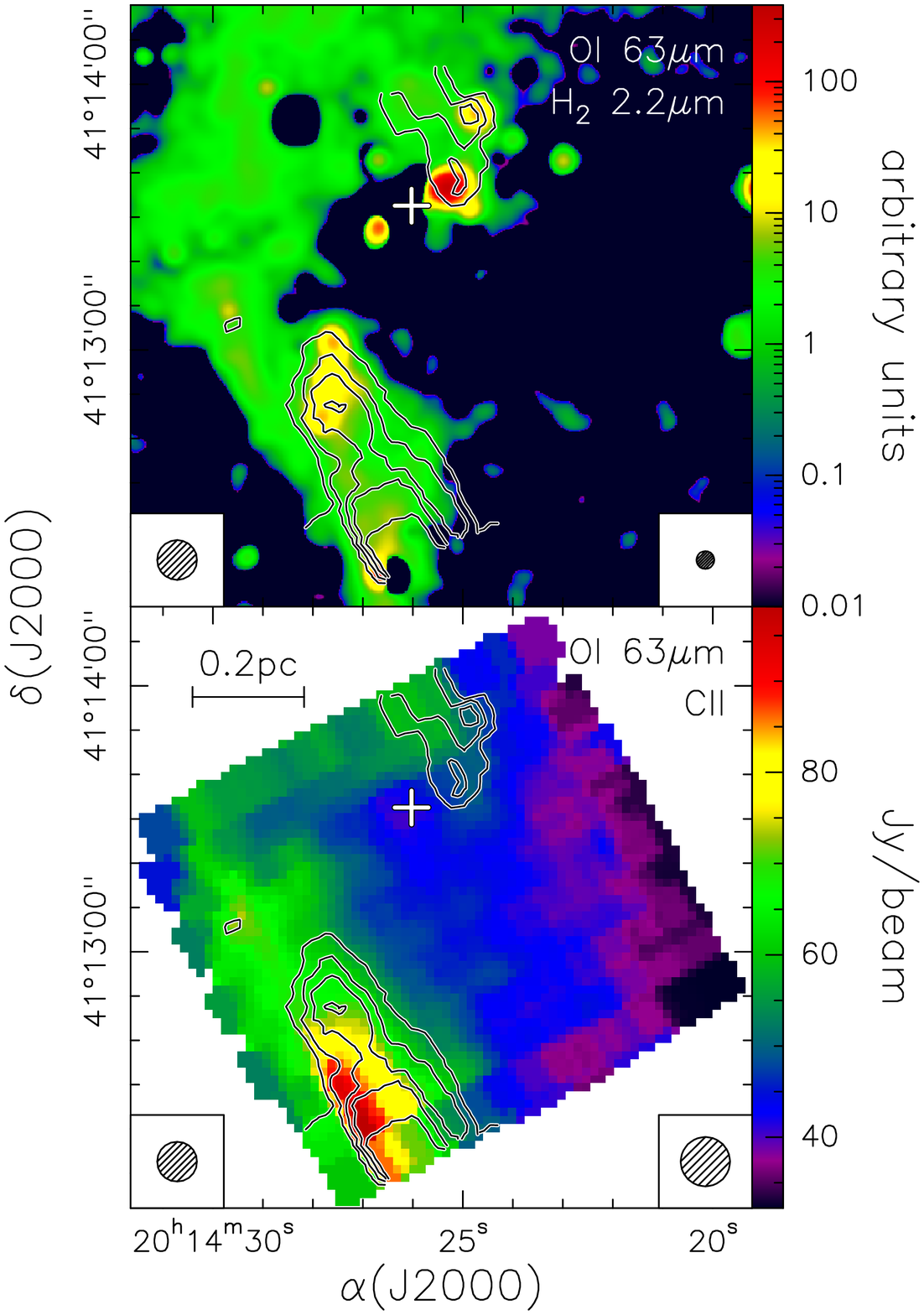}}
\caption{
Contour map of the \OI\ 63~\mic\ line overlaid on the
\HM\ 2.12~\mic\ line image from Cesaroni et al.~(\cite{cesa05}) (top panel)
and the \CII\ line map (bottom panel). The circles in the bottom left are the
HPBW of the \OI\ map, while those in the bottom right are the HPBWs of the other
lines. Contour levels range from 13.87 to 29.01 in steps of 3.78~Jy/beam.
}
\label{fmover}
\end{figure*}

In contrast to \CO\ and \WAT, the atomic lines could arise from the PDR
at the border of the clump (see Sect.~\ref{scon}) or the shocked region
along the southern lobe of the molecular outflow from \I.
To shed light on the nature of the atomic emission, in Fig.~\ref{fmover}
we overlay the \OI\ 63~\mic\ map on those of the \HM~2.12~\mic\ (from
Cesaroni et al.~\cite{cesa05}) and \CII\ lines.
Two facts are noteworthy. The first is that the peak of the \OI\ emission
does not coincide with that of the \CII\ emission. The second is that,
in contrast, the distribution of the \OI\ emission matches that of
the \HM\ knots well. Although \CII\ can also be excited in shocks and not
only in PDRs (Kristensen et al. ~\cite{krist13}; Benz et al.~\cite{benz16}),
these two lines of evidence indicate that in the case of \I\ the \OI\
line is in all likelihood associated with shocks in the outflow/jet,
while most of the \CII\ emission traces a PDR.
Finally, it is worth noting that unresolved \OI\ emission is also seen towards
the disk position (see Fig.~\ref{fmoch}), although only in the 145~\mic\
line (the 63~\mic\ is affected by an artefact at that position), which is consistent
with an analogous detection in disks around low-mass protostars (see
e.g. Fedele et al.~\cite{fed13}).

\section{Analysis and discussion}
\label{sdis}

\subsection{Luminosity}
\label{slum}

\begin{figure*}
\centering
\resizebox{16.5cm}{!}{\includegraphics[angle=0]{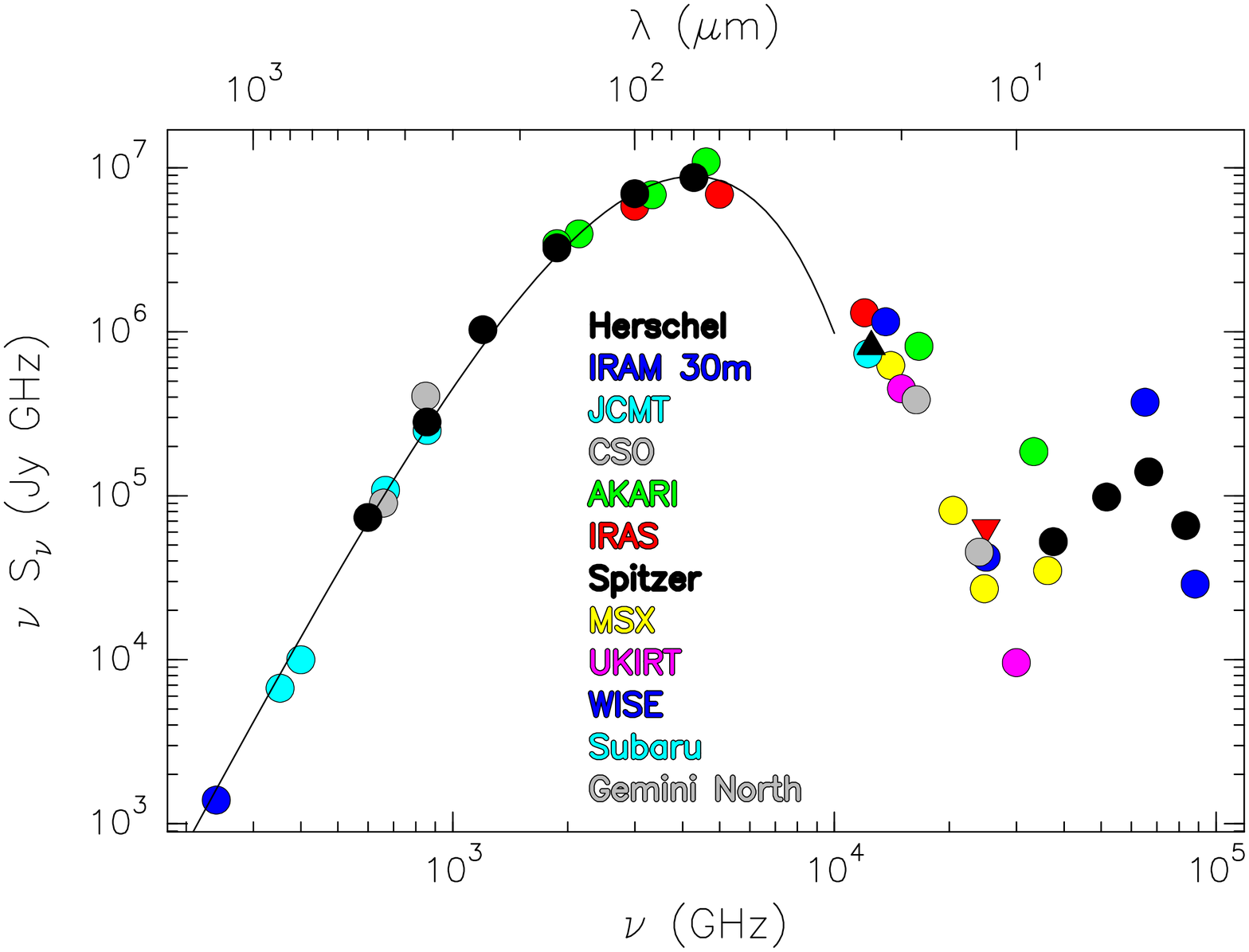}}
\caption{
Spectral energy distribution of \I\ obtained from the data in
Table~\ref{tsed}.  Triangles indicate lower and upper limits. The solid
curve is the best fit to the spectrum above 40~\mic\ obtained with the
model described in Sect.~\ref{smass}.
}
\label{fsed}
\end{figure*}

A reliable estimate of the luminosity of a YSO
relies on the availability of high-quality images of the continuum
emission at far-IR wavelengths, where the spectral energy distribution
(SED) of these objects peaks. In particular, the angular resolution
must be good enough to distinguish the emission associated with the YSO
from that possibly arising from unrelated nearby objects. As illustrated in
Sect.~\ref{scon}, our Herschel maps do indeed resolve the region around \I\ into
two components, one being quite compact and peaking at the disk position and the
other being more diffuse and approximately coincident with the PDR to the south-east. We
constructed the SED of the YSO(s) in \I\ by
considering only the contribution of the compact component. For this purpose,
we also used archival data and values from the literature, ranging
from the radio to the near-IR domain. In Table~\ref{tsed} we give the
flux densities measured at the different bands with references, and in
Fig.~\ref{fsed} we plot the corresponding SED. In most cases, the flux
densities were taken from catalogues (e.g. IRAS PSC), when available, or from
the literature; otherwise, the flux was estimated from the images directly.

\begin{table}
\caption[]{
Total flux densities of the continuum emission from \I.
}
\label{tsed}
\centering
\begin{tabular}{cccl}
\hline
\hline
$\lambda$ & $\nu$   & $S_\nu$     & Reference \\
 (\mic)   & (GHz)   & (Jy)       & \\
\hline
     3.4 & 88173.5  &   0.328    & WISE source catalogue \\
     3.6 & 83275.7  &   0.79     & Johnston et al. (\cite{johns}) \\
     4.5 & 66620.5  &   2.1      & Johnston et al. (\cite{johns}) \\
     5.8 & 51688.4  &   1.9      & Johnston et al. (\cite{johns}) \\
     4.6 & 65171.7  &   5.704    & WISE source catalogue \\
     8.0 & 51688.4  &   1.4      & Johnston et al. (\cite{johns}) \\
     8.3 & 36206.8  &   0.9646  & MSX catalogue 6 \\
     9.0 & 33310.0  &   5.578    & AKARI PSC \\
    10.0 & 29979.2  &   0.32     & UKIRT (Cesaroni et al. \cite{cesa99}) \\
    12.0 & 24982.5  &   1.687    & WISE source catalogue \\
    12.0 & 24982.7  &  $<$2.546  & IRAS PSC \\
    12.1 & 24715.0  &   1.0970   & MSX catalogue 6 \\
    12.5 & 23983.0  &   1.89     & Gemini North (De Buizer \cite{debui}) \\
    14.7 & 20463.5  &   3.9767   & MSX catalogue 6 \\
    18.0 & 16655.0  &  48.96     & AKARI PSC \\
    18.3 & 16382.0  &  23.5      & Gemini North (De Buizer \cite{debui}) \\
    20.0 & 14989.6  &  30        & UKIRT (Cesaroni et al. \cite{cesa99}) \\
    21.3 & 14048.4  &  44.321    & MSX catalogue 6 \\
    22.0 & 13626.8  &  84.482    & WISE source catalogue \\
    24.0 & 12491.0  & $>$65      & Spitzer \\ 
    24.5 & 12236.0  &  60        & Subaru (de Wit et al. \cite{dewit}) \\ 
    25.0 & 11991.7  & 108.9      & IRAS PSC \\
    60.0 &  4996.54 & 1381       & IRAS PSC \\
    65.0 &  4612.15 & 2358       & AKARI (from image) \\
    70.0 &  4283.00 & 2037       & Herschel (this work) \\
    90.0 &  3331.00 & 2057       & AKARI (from image) \\
   100.0 &  2997.92 & 1947       & IRAS PSC \\
   100.0 &  2998.00 & 2320       & Herschel (this work) \\
   140.0 &  2141.36 & 1848       & AKARI (from image) \\
   160.0 &  1873.69 & 1735       & Herschel (this work) \\
   160.0 &  1873.69 & 1840       & AKARI (from image) \\
   250.0 &  1199.00 &  859       & Herschel (this work) \\
   350.0 &   856.55 &  292       & JCMT (Cesaroni et al. \cite{cesa99}) \\
   350.2 &   856.00 &  329       & Herschel (this work) \\
   352.9 &   849.40 &  477       & CSO (Shinnaga et al. \cite{shinn}) \\ 
   450.0 &   666.20 &  162       & JCMT (Cesaroni et al. \cite{cesa99}) \\
   455.4 &   658.30 &  137       & CSO (Shinnaga et al. \cite{shinn}) \\ 
   500.5 &   599.00 &  123       & Herschel (this work) \\
   750.0 &   399.72 &   25       & JCMT (Cesaroni et al. \cite{cesa99}) \\
   850.0 &   352.70 &   19       & JCMT (Cesaroni et al. \cite{cesa99}) \\
  1249.1 &   240.00 &    5.8     & IRAM (Beuther et al. \cite{beu02a}) \\ 
\hline
\end{tabular}
\end{table}

\begin{figure}
\centering
\resizebox{8.5cm}{!}{\includegraphics[angle=0]{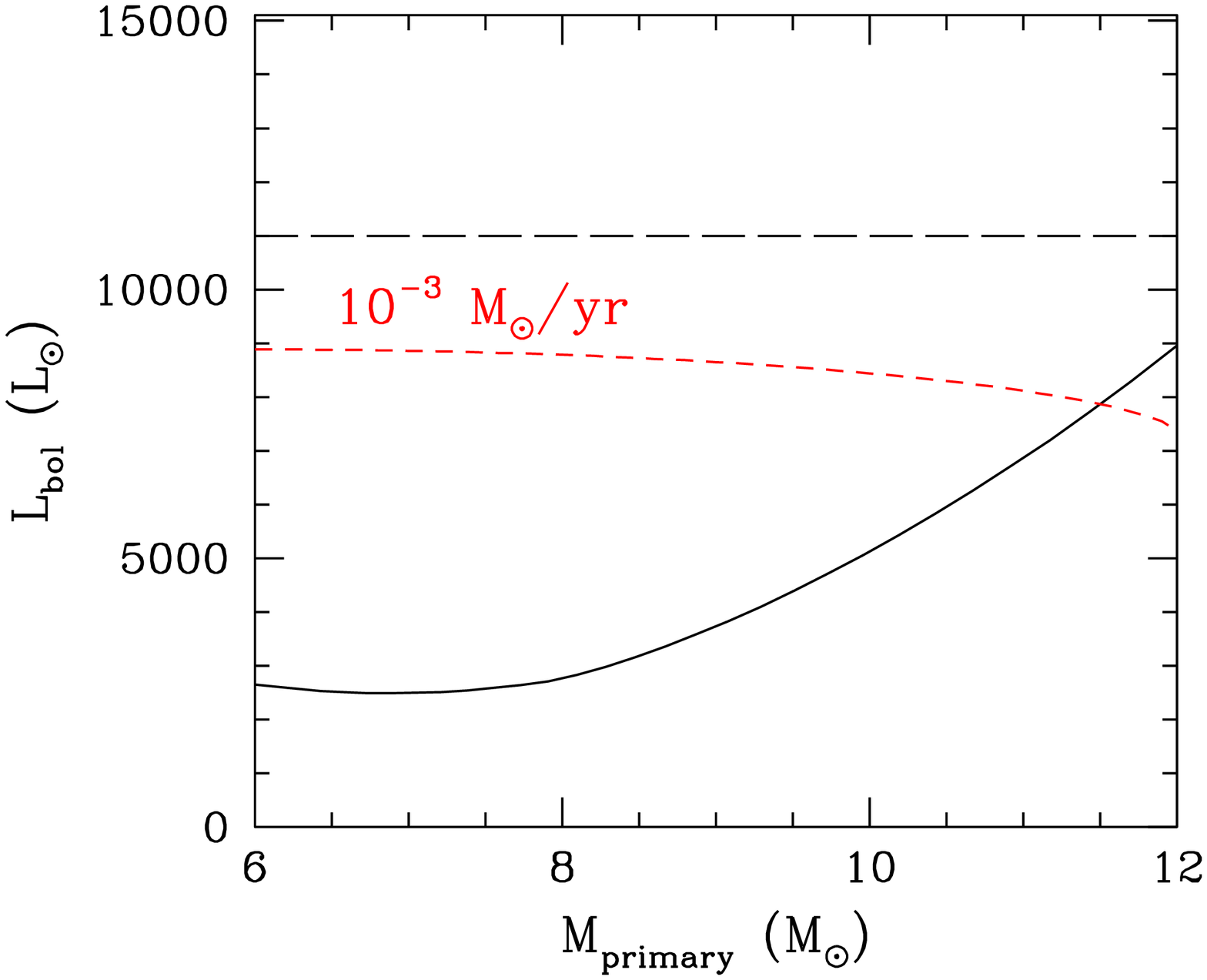}}
\caption{
Luminosity of a binary system with a total mass of 12~\Msun versus the mass of
the primary member. The solid line corresponds to two ZAMS stars. The red
dashed curve is for two protostars deriving their luminosity
from accretion, for an accretion rate of $10^{-3}~M_\odot$~yr$^{-1}$.
The long-dashed horizontal line marks the
bolometric luminosity of \I.
}
\label{fbin}
\end{figure}

As one can see from Fig.~\ref{fsed}, the bolometric luminosity is clearly
dominated by the emission around 100~\mic\ and can be estimated by integrating
the emission under the SED. For this purpose we have linearly interpolated
the flux densities in the $\Log\,S_\nu$--$\Log\,\nu$ plot and obtained $L_{\rm
bol}\simeq 1.1\times10^4$~\Lsun. Now that a precise estimate of the
luminosity is available, one may wonder if such a luminosity is dominated
by a single massive star or if it contains a significant contribution from other
nearby stars. In particular, it is of interest to establish if there is
a single star or a binary system at the centre of the disk. Since the
mass needed to reproduce the Keplerian pattern of the circumstellar disk
is 12~\Msun\ (see Chen et al.~\cite{chen}), in Fig.~\ref{fbin}, we have plotted
the total luminosity of a binary system of total mass 12~\Msun, as a
function of the mass of the primary member. This was computed for both two
zero-age main-sequence (ZAMS) stars and two protostars. For the latter, we
referred to Hosokawa et al.~(\cite{hoso09}), who demonstrate that the
protostellar luminosity up to $\sim$10~\Msun\ is dominated by the accretion
luminosity, whose approximate expression can be obtained from their Eqs.~(7)
and~(13), with the latter having originally been derived by Stahler et al.~(\cite{sps86}).

The solid curve in the figure clearly shows that the total luminosity
of two ZAMS stars drops dramatically as soon as the mass of the primary
decreases from the maximum value of 12~\Msun. In fact, even in concentrating
the whole mass in a single star, one can obtain only $\sim$80\% of the
measured bolometric luminosity. The remaining 20\% could be made up of nearby
lower-mass stars. In fact, according to the simulation described
by S\'anchez-Monge et al.~(\cite{sanch}), the most massive member of a
stellar cluster emitting $1.1\times10^4$~\Lsun is expected to be just a
$\sim$12~\Msun\ star.

In contrast, if the two components of the binary system are protostars,
for a fiducial accretion rate of $10^{-3}~M_\odot$~yr$^{-1}$ (see Cesaroni
et al.~\cite{cesa99}), the total luminosity is represented by the red dashed
line in Fig.~\ref{fbin}, which is rather insensitive to the way the mass
is partitioned between the primary and the secondary. In this case it seems
impossible to decide whether the mass of the system is mostly concentrated
in one of the members or more equally shared between them. However, we
note that the value of $10^{-3}~M_\odot$~yr$^{-1}$ is in all likelihood
the infall rate through the envelope enshrouding the embedded (proto)star
and not the accretion rate through the disk, and hence it must be
considered
as an upper limit. Indeed, detailed model fits to the SED (Johnston et
al.~\cite{johns}) and line emission (Chen et al.~\cite{chen}) derive
disk accretion rates $\la10^{-4}~M_\odot$~yr$^{-1}$. These models prove
that the luminosity of \I\ could originate from a $\sim$12~\Msun\ ZAMS
star with a non-negligible contribution from residual accretion, which is
also our conclusion.
It is also worth noting that the difference between the infall and
accretion rates, that is $\la$$10^{-3}~M_\odot$~yr$^{-1}$, could correspond
to the material ejected into the outflow. Indeed, the estimates of the
outflow mass-loss rate in \I\ range from $8\times10^{-4}~M_\odot$~yr$^{-1}$
(Shepherd et al.~\cite{shep}) to $1.6\times10^{-2}~M_\odot$~yr$^{-1}$
(Cesaroni et al.~\cite{cesa97}), which correspond to ejection rates
an order of magnitude lower for the internal jet entraining the outflow
(see e.g. Beuther et al.~\cite{beu02b}).
This implies ejection rates of about $10^{-4}$--$10^{-3}~M_\odot$~yr$^{-1}$,
which is consistent with the difference between infall and accretion rates.

\subsection{Temperature and density structure}
\label{stn}

\subsubsection{Derivation from continuum emission}
\label{stdcont}

\begin{figure*}
\centering
\resizebox{8.5cm}{!}{\includegraphics[angle=0]{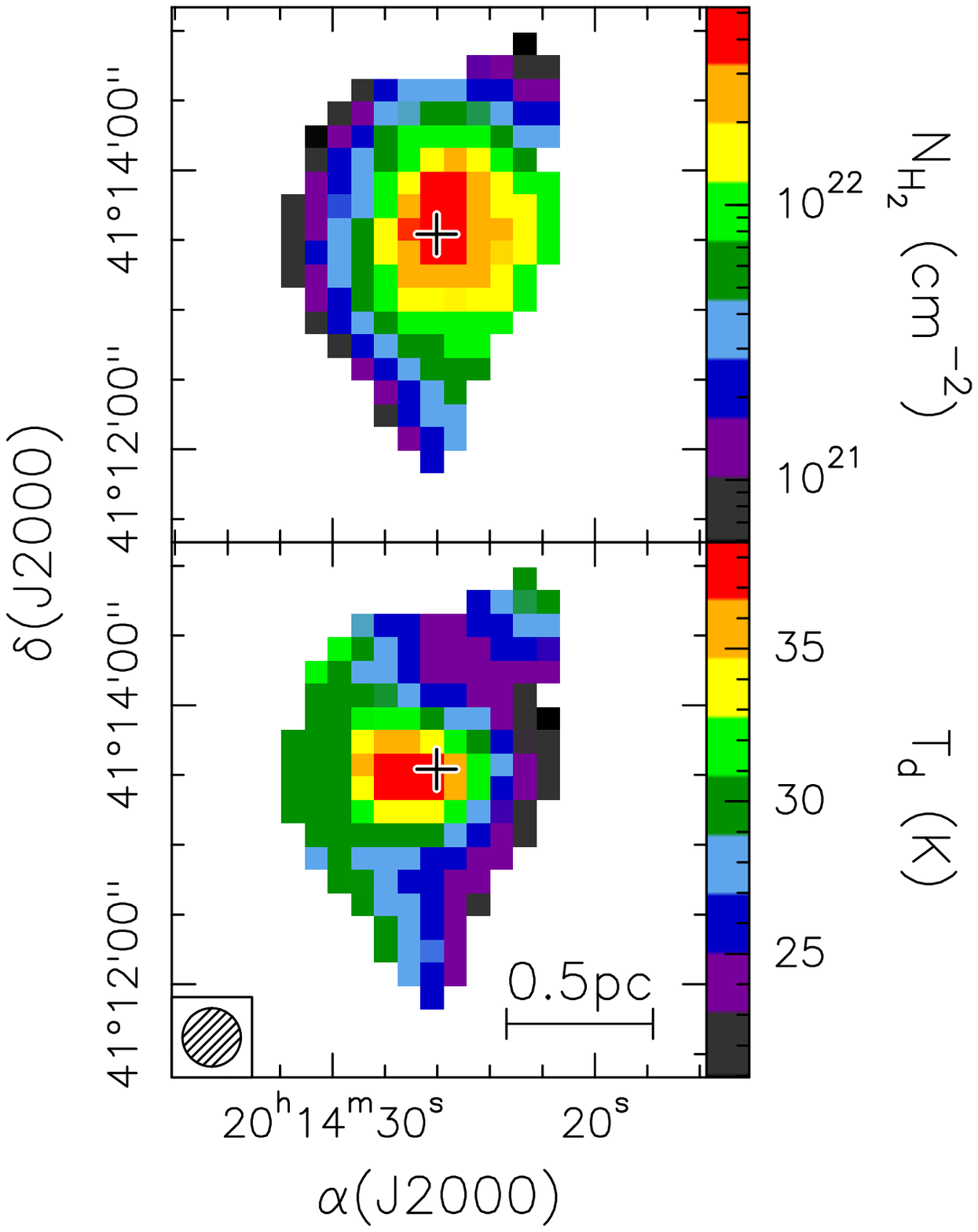}}
\hspace*{3mm}
\resizebox{8.5cm}{!}{\includegraphics[angle=0]{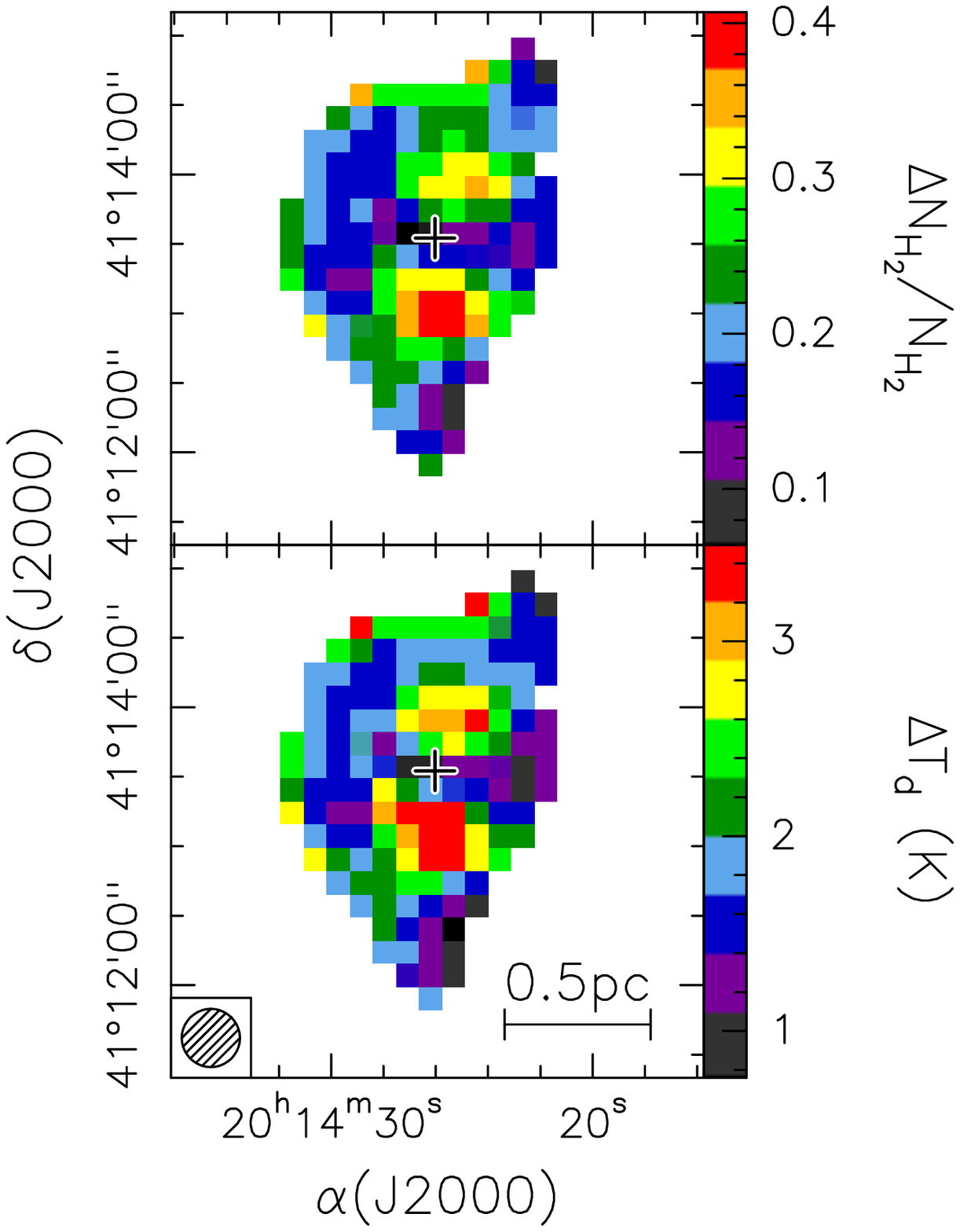}}
\caption{
Maps of the gas column density and dust temperature, assuming $\beta=1.5$ (left) and
corresponding errors (right). The angular resolution is represented by the
circle in the bottom left. The cross marks the position of the circumstellar
disk.
}
\label{fmtn}
\end{figure*}

Since we have obtained maps of the \I\ clump at six wavelengths across the
peak of the SED, we could derive a temperature and column density estimate all
over these maps by fitting the Herschel fluxes. In practice, we decided to
ignore the 500~\mic\ maps, whose HPBW ($\sim$36\arcsec) is the largest, and
thus achieved a compromise between the quality of the fit and angular resolution.
All maps were smoothed to the angular resolution of the
350~\mic\
map,
that is 25\arcsec,
and resampled on the same grid. Then a SED was extracted for each pixel,
rejecting the SEDs with one or more fluxes lying below the corresponding
3$\sigma$ level. Finally, by means of command MFIT of GILDAS, we fitted the SEDs
with
a modified black body using the expression
\begin{equation}
 S_\nu = \Omega_{\rm B} \, B_\nu(T)\left(1-\e^{-\tau}\right)  \label{esnu}
,\end{equation}
where $B_\nu$ is the Planck function, $\Omega_{\rm B}$ the solid angle of the beam,
$T$ the dust temperature, and $\tau$ the dust opacity. Moreover,
\begin{equation}
 \tau = \kappa_0 \left(\frac{\nu}{\nu_0}\right)^\beta \frac{\mu m_{\rm H}}{\mathcal{R}} N   \label{etau}
,\end{equation}
with $\kappa_0$ the dust absorption coefficient at frequency $\nu_0$, $\mu$
the mean molecular weight, $m_{\rm H}$ the mass of the hydrogen atom, $\mathcal{R}$
the gas-to-dust mass ratio, and $N$ the column density along the line of sight. For
our calculation we adopted the values $\kappa_0=1$~cm$^2$~g$^{-1}$,
$\nu_0=230$~GHz, and $\mu=2.8$, $\mathcal{R}=100$.

The free parameters of the fit are the dust temperature and column density,
while $\beta$ is assumed equal to the value of 1.5, obtained by fitting
the global SED in Fig.~\ref{fsed} above 40~\mic\ with Eq.~(\ref{esnu}),
where $\Omega_{\rm B}$ is replaced by the solid angle subtended by the
whole clump, $\Omc=\pi\theta^2$, with $\theta\simeq70\arcsec$
(see Fig.~\ref{fmcon}). The same fit also provides us with an estimate for
the mass and temperature of the clump 150~\Msun\ and 33~K, respectively
(but see Sect.~\ref{smass}).

In Fig.~\ref{fmtn} we show the resulting maps and the corresponding
errors. As expected, the column density and temperature peak approximately
at the position of the disk, within the angular resolution of the images.
The column density map allowed us to derive a more precise estimate of the
clump mass, as done later in Sect.~\ref{smass}.

\begin{figure}
\centering
\resizebox{8.5cm}{!}{\includegraphics[angle=0]{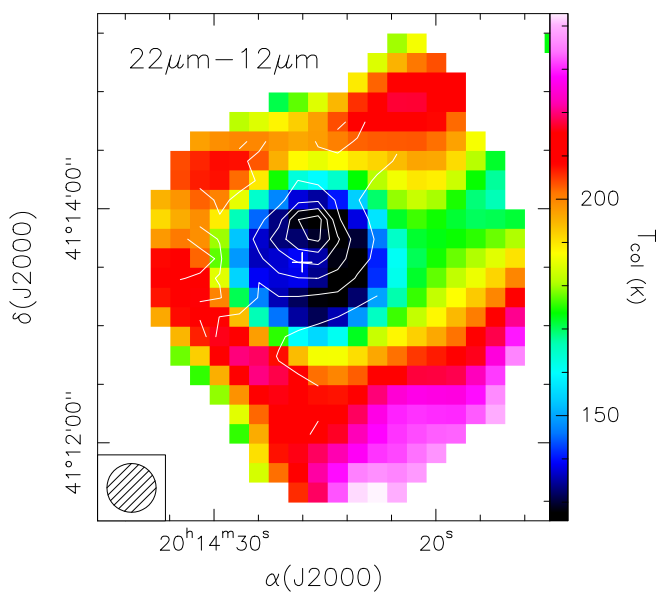}}
\caption{Map of the dust temperature from Fig.~\ref{fmtn} overlaid on the
map of the colour temperature obtained from the WISE 12 and 22~\mic\
images. The angular resolution is denoted by the circle in the bottom left.
}
\label{fmtc}
\end{figure}

All the above holds for the cold-dust component, but it is also interesting to
explore the temperature distribution of the hot-dust component contributing
to the continuum emission below $\sim$40~\mic. This can be seen in a map
of the colour temperature obtained from the WISE 22 and 12~\mic\ images,
after smoothing to the same angular resolution. The result is shown in
Fig.~\ref{fmtc}, where we have also overlaid the temperature map of the cold-dust
component (from Fig.~\ref{fmtn}), for the sake of comparison. Clearly,
the dust temperature peaks where the colour temperature experiences a dip,
which is consistent with
most of the near-IR emission being confined to the surface of the clump.

\subsubsection{Derivation from \CO\ emission}
\label{scoem}

\begin{figure}
\centering
\resizebox{8.5cm}{!}{\includegraphics[angle=0]{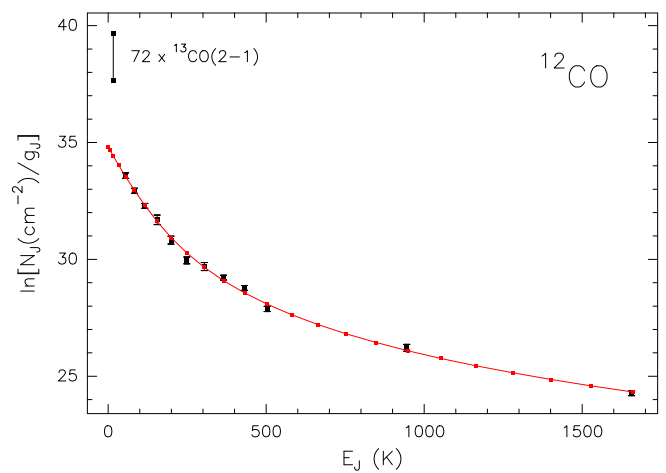}}
\caption{
Rotation diagram of \CO. The black points with error bars represent the
Herschel data, whereas the two black points connected by a bar are the lower
and upper limit to the \COI\ column density in the $J=2$ level obtained
from the \COI(2--1) data of Cesaroni et al.~(\cite{cfw99}) after multiplying
the column density for a \CO/\COI\ abundance ratio of 72. The red points
connected by the curve are the best fit to the Herschel data obtained with
the model described in Appendix~\ref{sapp}.
}
\label{frotco}
\end{figure}

Another description of the temperature and density distribution in
the clump enshrouding \I\ could be obtained from the \CO\ emission.
In Fig.~\ref{frotco}, we show the rotation diagram obtained from the
\CO\ emission integrated over the line and the whole emitting area, whose
angular size is $\sim$150\arcsec\ (see Fig.~\ref{fmco}).
One can see that the points are not distributed along a straight line,
which is expected if the source were isothermal and homogeneous. This is not
surprising because we know from our analysis in Sect~\ref{stdcont} that
the temperature varies over the clump. To fit the rotation diagram,
one thus needs a model that takes temperature and, possibly,
density gradients into account. For this purpose, we referred to the approach adopted by
Mauersberger et al.~(\cite{mau88}; hereafter MWH -- see their appendix)
and Neufeld~(\cite{neuf12}).
Our
derivation of the mean column density, $N_J$, in a rotational level of \CO\
is slightly different from that of those authors and is described in
Appendix~\ref{sapp}. In practice, we express $N_J$ as a function of the
energy of the level, $E_J$, through a number of variables that are all
known, except the temperature at the surface of the clump, $\To$, the mean
column density of \CO, $\NCO$, and the index $\alpha=(q-p-3)/q$, where $q$
and $p$ are the power-law exponents of the temperature and
density profiles, respectively.

The best fit to the rotation diagram is shown by the red curve in
Fig.~\ref{frotco} and corresponds to the minimum $\chi^2$ obtained by varying
the three free parameters over suitable ranges. We thus obtain
$\To=32\pm5$~K,
$\NCO=(2.2^{+1.8}_{-1.0})\times10^{16}$cm$^{-2}$, and
$\alpha=3.15\pm0.13$.
We note that all the above assumes local thermodynamic equilibrium (LTE) and optically thin \CO\ lines.
To verify if
these assumptions are valid,
we
ran the programme RADEX by Van der Tak et al.~(\cite{vdtak}) using -- as inputs -- the
column density obtained from the fit in Fig.~\ref{frotco},
a temperature
and \HM\ density spanning the ranges 30--200~K and
$10^3$--$10^6$~\cmc,  respectively, and a line width of
8~\kms\ (see Fig.~1 of Kawamura et al.~\cite{kawa99}). The results confirm
that all lines are optically thin, with the highest value of the opacity,
0.3, being obtained for the \CO(4--3) transition. We conclude that our approach
is plausible and self-consistent.

The mass of the clump, obtained from Eq.~(\ref{emass}), is
$M_{\rm H_2} \simeq 5.5^{+4.6}_{-2.5}~M_\odot$
for an abundance of \CO\ relative to H$_2$ of $10^{-4}$.
The value of $M_{\rm H_2}$ is by far too small compared to other estimates
such as that obtained from the continuum emission (see Sect.~\ref{stdcont})
or the one derived by Shepherd et al. (\cite{shep}) from their \CO\ and
\COI\ (1--0) interferometric maps ($\sim$300~\Msun). Such a discrepancy
indicates that most of the material must be cold enough to be traced only by
the low-energy lines of \CO\ and its isotopologues. To include the contribution
of these lines in the rotation diagram, we have used the single-dish \COI(2--1)
data of Cesaroni et al.~(\cite{cfw99}). These authors mapped only a
limited portion of the whole clump, about 48\arcsec$\times$48\arcsec\
centred around the peak of the emission. Therefore, we could derive an upper
limit of $1.17\times10^{16}$~\cmq\ to the mean column density over $\Omc$
in the $J=2$ level of \COI, by assuming that the part of the clump that was
not mapped in \COI(2--1) has the same mean column density as the mapped
one. Conversely, a lower limit of $1.56\times10^{15}$~\cmq\ was obtained
by assuming that no \COI\ emission is present beyond the region mapped by
Cesaroni et al.~(\cite{cfw99}).

In Fig.~\ref{frotco}, we have plotted the corresponding lower and upper limits
on the \CO\ column density obtained by multiplying the \COI\ values by
a ratio between the \CO\ and \COI\ abundances of 72, estimated from
Wilson \& Rood~(\cite{wilroo}) for a galactocentric distance of 8.2~kpc.
From the rotation diagram of the (2--1) and (4--3) lines only, one obtains
a temperature and \CO\ column density in the range 6.4--9.6
K and $4.2\times10^{17}$--$4.8\times10^{18}$~\cmq, respectively. The latter corresponds
to $M_{\rm H_2}$=90--1050~\Msun, a range consistent with the estimates obtained
in other tracers. This confirms that the majority of the clump is sufficently
cold to be seen only in the low-energy transitions of the CO isotopologues.

Despite their limited utility in sampling the cold outer region of the clump,
where most of the mass is concentrated, the high-energy transitions of \CO\
still provide us with important information on the innermost structure of the
clump. In Sect.~\ref{scoem}, we have derived $\alpha=(q-p-3)/q\simeq3.15$. It
is also known that $q$ is related to the spectral index $\beta$ of
the optically thin (sub-)millimetre continuum emission by the relation
$q=-2/(4+\beta)$ (see e.g. Doty \& Leung~\cite{dole}).  Since in our case
$\beta\simeq1.5$ (see Sect.~\ref{stdcont}), we obtain $q\simeq-0.37$ and,
consequently, $p\simeq-2.2$, which proves that the density distribution
inside the clump is steeply peaked towards the centre.  In all likelihood,
this indicates that the envelope enshrouding \I\ is currently collapsing
and accreting onto the disk.

\subsection{Mass of the clump}
\label{smass}

The results obtained so far can be used to derive an accurate estimate of the
clump mass. We have already mentioned in Sect.~\ref{stdcont}
that a mass estimate can be obtained from a fit to the (sub-)millimetre and far-IR
part of the SED, using a modified black body that assumes the clump to
be isothermal and homogeneous. The best fit provides us with a mass
of 150~\Msun\ and a temperature of 33~K. However, as demonstrated in
Sect.~\ref{stn}, there are significant temperature and density variations
around \I, so that a more realistic model is needed to improve upon these
estimates.

An estimate taking such temperature and density variations into account
can be directly obtained from the map in Fig.~\ref{fmtn}, by
integrating the column density over the whole clump:
\begin{equation}
\Mc = \mu m_{\rm H} \sum_i N_i ~ \Delta\Omega \, d^2
,\end{equation}
where $i$ indicates a pixel of the column density map in Fig.~\ref{fmtn},
$N_i$ is the gas column density at pixel $i$, $\Delta\Omega=100$~arcsec$^2$ is
the pixel's solid angle, the sum extends over all pixels where the column
density has been estimated, $d$=1.64~kpc is the distance of \I, and the
other symbols have the same meaning as in Eq.~(\ref{etau}). We calculated
a total mass of 245~\Msun, which is significantly greater than the 150~\Msun\
estimated from the modified black-body fit.

Another way to derive $\Mc$ is to fit the (sub-)millimetre and far-IR part of the SED
with the model by Cesaroni~(\cite{rcesa19}), which takes both
temperature and density gradients into account in the form of power laws. For the dust
absorption coefficient, we used the same value adopted for Eq.~(\ref{etau}),
with $\beta=1.5$. The input parameters of the model are the radius of the
clump, $\Ro$, the ratio between the inner and outer radius, $\Ri/\Ro$,
the distance to the source, $d$, the power-law indices of the temperature,
$q$, and density, $p$, the temperature at the outer radius, $\To$, and the
mass of the clump, $\Mc$. From Sect.~\ref{scoem}, we know that $q=-0.37$
and $p=-2.2$, while from the maps in Fig.~\ref{fmcon} we measured an angular
radius of the clump of $\sim$70\arcsec, which implies $\Ro=0.56$~pc for
$d=1.64$~kpc. The minimum radius, $\Ri$, beyond which the spherically
symmetric approximation holds can be assumed to be the centrifugal radius of
the disk. The latter was estimated to be a few $\sim$1000~au, depending on
the model (Keto \& Zhang~\cite{ketzha}; Johnston et al.~\cite{johns}; Chen
et al.~\cite{chen}). Therefore, for our model fit, we assume $\Ri/\Ro=0.05$.

\begin{figure}
\centering
\resizebox{8.5cm}{!}{\includegraphics[angle=0]{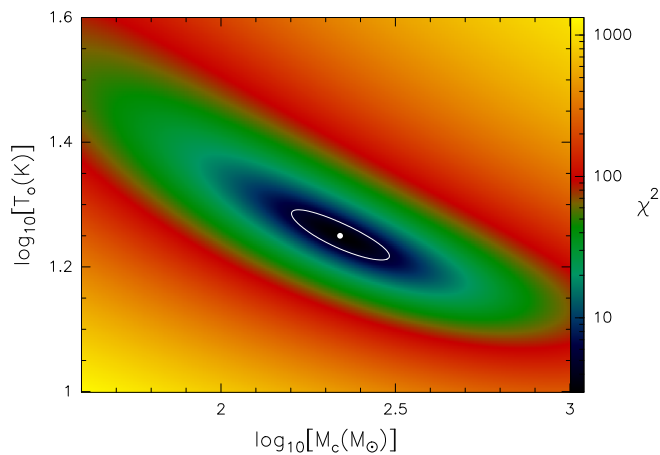}}
\caption{
Plot of the $\chi^2$ obtained by fitting the clump model of
Cesaroni~(\cite{rcesa19}) to the SED of \I\ (shown in Fig.~\ref{fsed}),
as a function of the clump surface temperature and mass.
For the $\chi^2$ calculation, we have assumed an error of 20\% for all fluxes.
The white dot marks the minimum corresponding to the best fit, while
the white contour is the 1$\sigma$ confidence level (i.e. 68\% reliability)
according to the criterion of Lampton et al.~(\cite{lamp}).
}
\label{fchiq}
\end{figure}

The
best fit to the SED for $\lambda>40$~\mic\ was obtained by varying $\Mc$
and $\To$ over suitable ranges. In Fig.~\ref{fchiq}, we have plotted the value
of $\chi^2$ as a function of these two variables, with the contour
corresponding to a confidence level of 1$\sigma$, according to the criterion
of Lampton et al.~(\cite{lamp}). In this calculation, we have assumed an
error of 20\% for all fluxes. We conclude that $\To=18^{+1.8}_{-1.5}$~K
and $\Mc=220^{+85}_{-60}$~\Msun. The latter value is in agreement with that
obtained by integrating the column density over the region, while the
surface temperature of the clump is comparable to the minimum value (21~K)
of the temperature map in Fig.~\ref{fmtn}.
We note that this value of the temperature is very similar to that expected
at a distance $\Ro=0.56$~pc from a star of $L_\ast=1.1\times10^4$~\Lsun. In
fact, following Zhang et al.~(\cite{zha02}), one obtains\footnote{We note that
the expression given by Zhang et al.~(\cite{zha02}) contains a typo: the
exponent $1/(1+\beta)$ must be replaced by $1/(4+\beta)$.}
\begin{equation}
\To = 65 \left(\frac{0.1{\rm pc}}{\Ro}\right)^\frac{2}{4+\beta}
         \left(1.25\frac{L_\ast}{10^5~\Lsun}\right)^\frac{1}{4+\beta}
         \simeq 24~K,
\end{equation}
with $\beta=1.5$ (see Sect.~\ref{stdcont}).

We conclude that our analysis indicates that the clump enshrouding \I\ has
a mass of $\sim$250~\Msun\ with a density distribution $\propto R^{-2.2}$,
while the temperature is relatively low, $\sim$20~K, at the surface of
the clump and increases as $R^{-0.37}$ towards the centre.

\subsection{Shocked region}
\label{sshock}

As noted in Sect.~\ref{slin}, the oxygen line emission very likely
originates from shocked gas, given the positional coincidence with \HM\
knots in the jet from \I. Under this hypothesis, one can derive an estimate
of the oxygen column density, \No, and \HM\ volume density, \nHM, from the
two \OI\ lines observed. For our calculations we used the programme RADEX (Van
der Tak et al.~\cite{vdtak}) with the oxygen atomic data from Sch\"oier et
al.~(\cite{shoier}), assuming collisions only with \HM\ molecules, a kinetic
temperature of 500~K, and a line width of 20~\kms,
a fiducial value suggested by observations of the oxygen line in other objects
(Kristensen et al.~\cite{krist17}; Yang et al.~\cite{yang22}).
The
results are weakly dependent on the temperature (see also Fig.~10 of Nisini
et al.~\cite{nisi}) and line width; therefore, our choice cannot significantly
affect the outcome of RADEX. We also note that if we were to choose collisions
with H instead of \HM, no solution would be found for \No\ or \nHM (see below).

\begin{figure}
\centering
\resizebox{8.5cm}{!}{\includegraphics[angle=0]{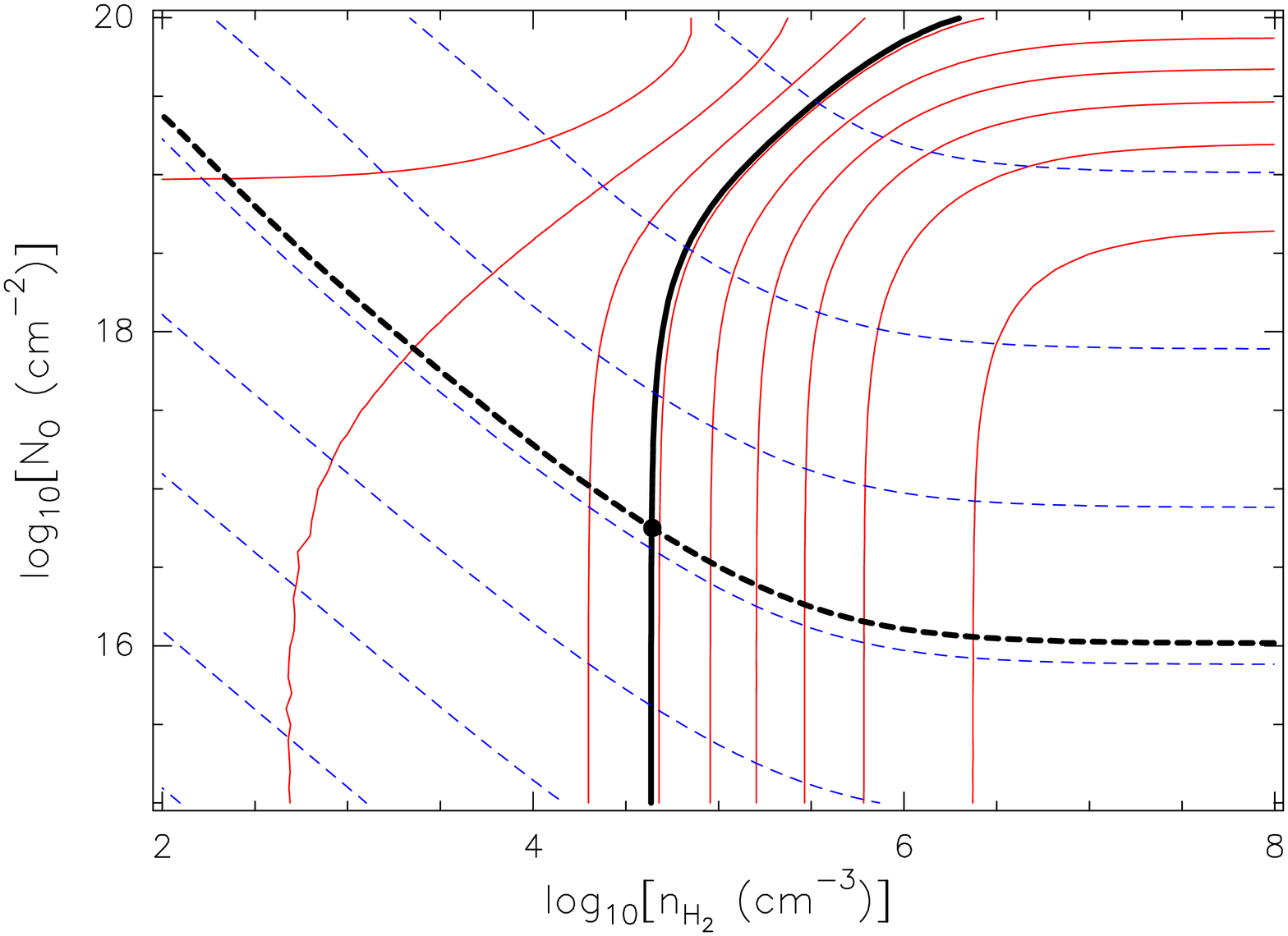}}
\caption{
Example of the method used to estimate the oxygen column density and
the molecular hydrogen volume density over the \OI-emitting region.
The solid red contours and the blue dashed contours are maps
of $\int T_{\rm 63\mu m} \d V / \int T_{\rm 145\mu m} \d V$ and
$\int T_{\rm 63\mu m} \d V$, respectively. The solid, thick black contour and the black dashed contour
correspond to the observed values of the same quantities. The dot marks
the intersection between these two contours, which gives the pair of
values of \nHM\ and \No\ ($4.4\times10^4$~\cmc\ and $5.6\times10^{16}$~\cmq)
that reproduce both observables.
}
\label{fexa}
\end{figure}

We searched for the pair of values \No,\nHM\ that can reproduce both the
ratio between the 63 and 145~\mic\ lines and the value of the 63~\mic\
line intensity. For this purpose, we computed the ratio between the maps
of the integrated emission in these two lines, after suitably resampling
and smoothing to the same resolution. Then, at each pixel, we extracted the
value of the line ratio and that of the 63~\mic\ brightness temperature
integrated over the line. These were used to derive \No\ and \nHM\ as
illustrated in Fig.~\ref{fexa}. In practice, we computed a grid of line
intensities (in K~\kms) for a wide range of \No\ and \nHM\ and then in the
plane \nHM,\No\ we plotted the contour levels corresponding to the observed
values of $\int T_{\rm 63\mu m} \d V$ and $\int T_{\rm 63\mu m} \d V / \int
T_{\rm 145\mu m} \d V$. The intersection between the two contours gives
the pair \nHM,\No\,, satisfying both conditions. As previously mentioned,
if collisions with only atomic hydrogen are included, the two contours do
not intersect and no solution is found.

\begin{figure}
\centering
\resizebox{8.5cm}{!}{\includegraphics[angle=0]{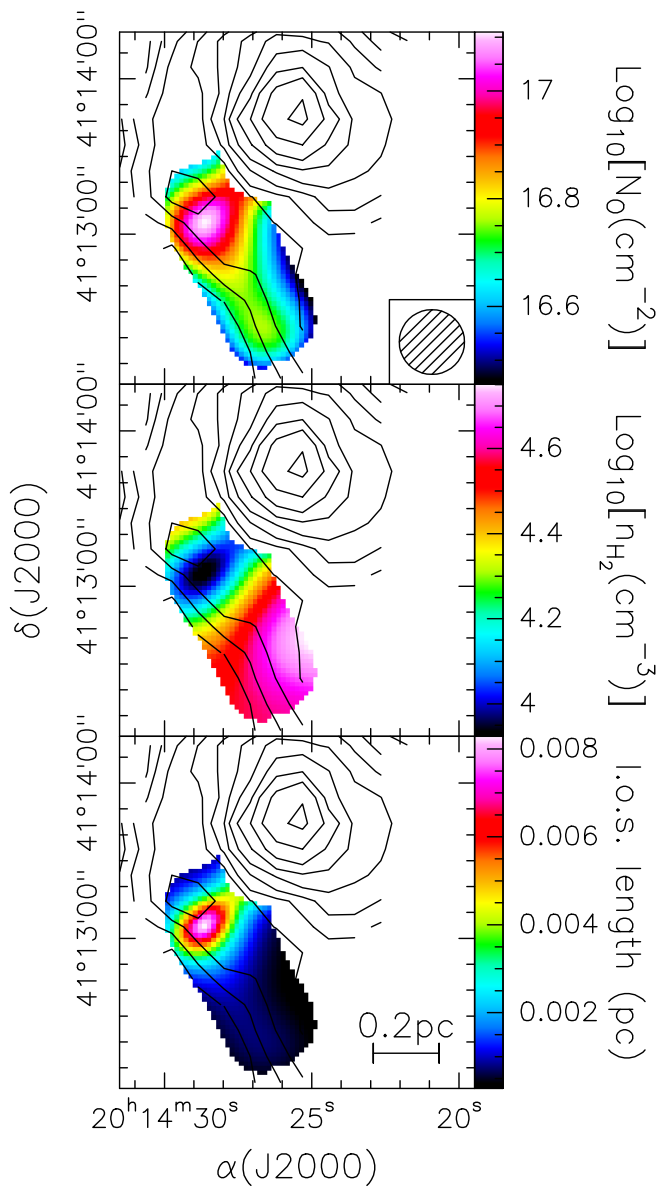}}
\caption{
Maps of the oxygen column density (top panel) and molecular hydrogen
volume density (middle panel) over the shocked region where \OI\ emission
is detected. The bottom panel shows a map of the geometrical thickness
along the line of sight of the same region, derived from the ratio between
\No\ and \nHM, assuming a constant oxygen abundance relative to
molecular hydrogen of $6\times10^{-4}$.
The contours represent the same map of the gas column density
as in Fig.~\ref{fmtn}.
}
\label{foxy}
\end{figure}

After applying this procedure to each pixel of the maps, we obtained the
distribution of \No\ and \nHM\ along the shocked region to the south-east of the
clump, where the southern lobe of the (precessing) jet from \I\ is seen.
The results are shown in Fig.~\ref{foxy}, where also a contour map of the
gas column density from Fig.~\ref{fmtn} has been overlaid for the sake of comparison.

From the ratio between \No\ and \nHM\,, one can obtain an estimate of the
thickness of the oxygen-emitting region along the line of sight. For
this purpose, we have assumed an abundance of oxygen with respect
to H equal to the typical interstellar-medium value of $3\times10^{-4}$ (Savage \&
Sembach~\cite{savsem}), which implies an abundance relative to \HM\
of $6\times10^{-4}$. The resulting map is shown in the bottom panel of
Fig.~\ref{foxy}. The result depends on the oxygen abundance, which could be
an order of magnitude lower than the value adopted by us (see e.g. Liseau \&
Justtanont~\cite{lis09}). However, even if the abundance were overestimated
by a factor $\sim$10, the geometrical thickness of the shocked region
would be much less than the size of the same region in the plane of the
sky. This suggests that we are looking at the shocked layer approximately
face on.

All the previous results favour an origin of the
\OI\ remission from dissociative shocks in the southern lobe of the jet powered
by \I. With this in mind, we used Eq.~(7) of Hollenbach~(\cite{holl}) to derive
the mass-loss rate in the jet, $\dot{M}_{\rm o}$, from the luminosity of the
\OI\ 63~\mic\ line. This equation can be used provided the density of the
shocked region and the shock velocity satisfy the condition $\nHM\,\varv_{\rm
sh}\la10^{12}$~cm$^{-2}$~s$^{-1}$. In our case, $\nHM<5\times10^4$~\cmc, as
shown in Fig.~\ref{foxy}, and $\varv_{\rm sh}\la100$~\kms, which was obtained
from proper motion measurements of the \WAT\ masers (Moscadelli et
al.~\cite{mosca05,mosca11}) and \HM\ knots (Massi et al.~\cite{massi23}).
Therefore, $\nHM\,\varv_{\rm sh} < 5\times10^{11}$cm$^{-2}$~s$^{-1}$,
which satisfies the previous condition.

The 63~\mic\ line luminosity was obtained by integrating over the
whole emitting region and has turned out to be $\sim$2~\Lsun, which implies
$\dot{M}_{\rm o}\simeq 4\times10^{-4}$~\Msun~yr$^{-1}$, where the result
has been multiplied by a factor 2 to take into account the fact that we have
analysed only one of the two lobes of the outflow. This
value is consistent within the uncertainties with that of
$8\times10^{-4}$~\Msun~yr$^{-1}$ obtained by Shepherd et al.~(\cite{shep})
from their \CO(1--0) high-resolution map and it is also the mass outflow rate
expected for a high-mass YSO of $\sim1.1\times10^4$~\Lsun, as one can see in
Fig.~4 of L\'opez-Sepulcre et al.~(\cite{lose09}) and Fig.~5 of Maude et
al.~(\cite{maud15}).
Despite the good agreement between the outflow rates obtained from \OI\ and
\CO, one must keep in mind that the two tracers are not necessarily related
for a variety of reasons, which has been extensively discussed by Mottram et al.~(\cite{mott17}).
In particular, the \OI\ line is probably associated with the current
accretion-ejection episode, whereas the
low-J \CO\ lines are
more representative of
the time-averaged accretion-ejection phenomenon. One could speculate that
the similarity between the two estimates in \I\ suggests that, in this
object, accretion proceeds in a less episodic way than in other sources where
significant accretion outbursts have been detected (Caratti o Garatti et
al.~\cite{cagana}; Hunter et al.~\cite{hunt}).

\section{Summary and conclusions}
\label{ssum}

Herschel observations of the continuum and line emission from the high-mass
(proto)star \I\ were performed with the main goal of deriving an accurate
estimate of the stellar luminosity, clump mass, and outflow mass-loss
rate. In particular, we imaged a region of $\sim$1~pc at six continuum
bands, in twelve \CO\ rotational transitions, and three atomic (\OI\
and \CII) lines. Serendipitously, also an \WAT\ line was detected.

From the continuum data, we have estimated a bolometric luminosity of
$1.1\times10^4$~\Lsun. We show that in all likelihood, this
luminosity is mostly contributed to by a $\sim$12~\Msun\ ZAMS star at the
centre of the Keplerian disk imaged in previous studies.

We have also derived maps of the gas column density and dust temperature.
Not surprisingly,
both the temperature and column density peak at the positions of the
disk. The mass of the clump has been obtained both by integrating the
column density over the mapped region and through a model fit to the
SED that takes temperature and density gradients  into account.
We have found that the clump mass is $\sim$250~\Msun\ with a steep
density distribution $\propto R^{-2.2}$, which supports the idea that
the envelope enshrouding the disk is infalling.

While the molecular transitions trace the majority of the clump seen in the
(sub-)millimetre continuum, the atomic lines are detected only towards the
disk and
south-eastern border of the clump. However,
the peak of the \OI\ emission is clearly offset from that of the \CII\
emission,
with the former coinciding with the jet knots revealed by the \HM\
emission at 2.12~\mic\ and the latter probably being associated with a PDR seen
at near-IR wavelengths.
This indicates that the \OI\ emission most likely originates from shocks.
At the same time, one cannot rule out that part of the \CII\ line
emission also comes from shocks, as some \CII\ emission is seen towards the
peak of \OI.
From the maps of the two \OI\ lines, we have determined the
distribution of the oxygen column density and \HM\ volume density over
the shocked border by fitting the data with the RADEX numerical code.
Finally, the \OI\ emission arising from dissociative shocks has been
used to derive an estimate of the outflow mass-loss rate, which is
$\sim$$4\times10^{-4}$~\Msun~yr$^{-1}$. Such a value is indeed expected
for a YSO of $1.1\times10^4$~\Lsun\
(L\'opez-Sepulcre et al.~\cite{lose09}; Maude et al.~\cite{maud15}).

\begin{acknowledgements}
We thank the anymous referee for a careful reading of the manuscript
and constructive critcisms.
This work was partly supported by the Italian Ministero dell’Istruzione,
Universit\`{a} e Ricerca through the grant Progetti Premiali 2012-iALMA
(CUP C52I13000140001). This project has received funding from the European
Union’s Horizon 2020 research and innovation programme under the Marie
Sklodowska- Curie grant agreement No 823823 (DUSTBUSTERS) and from the
European Research Council (ERC) via the ERC Synergy Grant ECOGAL (grant
855130).
{\it Herschel} is an ESA space observatory with science
instruments provided by European-led Principal Investigator consortia and
with important participation from NASA. PACS has been developed by a
consortium of institutes led by MPE (Germany) and including UVIE (Austria);
KUL, CSL, IMEC (Belgium); CEA, OAM P (France); MPIA (Germany); IAPS, OAP/OAT,
OAA/CAISMI, LENS, SISSA (Italy); IAC (Spain). This development has been
supported by the funding agencies BMVIT (Austria), ESA-PRODEX (Belgium),
CEA/CNES (France), DLR (Germany), ASI (Italy), and CICYT/MCYT (Spain).
SPIRE has been developed by a consortium of institutes led by Cardiff Univ. (UK)
and including Univ. Lethbridge (Canada); NAOC (China); CEA, LAM (France);
IAPS, Univ. Padua (Italy); IAC (Spain); Stockholm Observatory (Sweden);
Imperial College London, RAL, UCL-MSSL, UKATC, Univ. Sussex (UK); Caltech,
JPL, NHSC, Univ. Colorado (USA). This development has been supported by
national funding agencies: CSA (Canada); NA OC (China); CEA, CNES, CNRS
(France); ASI (Italy); MCINN (Spain); Stockholm Observatory (Sweden); STFC
(UK); and NASA (USA).
\end{acknowledgements}

\begin{appendix}

\section{Rotation diagram with temperature and density gradients}
\label{sapp}

We present a model to fit the rotation diagram of a molecule, which
takes
the presence of temperature and density gradients  into account.
Our approach is inspired by the modified Boltzmann relation of MWH
and Neufeld~(\cite{neuf12}), but it differs from their approaches for a number
of reasons. The Neufeld method is similar to ours, but it does not take density gradients  into account nor does it provide an expression for the
CO column density -- as we illustrate below with Eq.~(\ref{enco}). As for MWH, they
assume an infinite clump, whereas we consider a clump of finite radius.
More specifically,
assuming spherical symmetry and temperature and density
laws for the molecule of the type $T\propto R^q$ and $n\propto R^p$, with
$R$ being the distance from the centre, MWH found that the mean column density
over the cloud in a level of energy $E$, divided by the statistical weight,
is proportional to $E^{-\alpha}$, where $\alpha$ is a function of $q$ and
$p$. They concluded that the relationship between column density and energy
should be fitted by a straight line in a modified rotation diagram where the
logarithms of both quantities are reported (in contrast to the traditional
rotation diagram where the linear relationship is expected in a plot of
$\log N$ vs $E$). The problem with this approach is that it works only if
the level energy is sufficiently high. In particular, it fails to describe
the ground state level, where $E=0$ and the expression for the column
density diverges. Moreover, the expression obtained by MWH
allows derivation of $\alpha$, but it cannot be used to
compute the temperature and total column density of the molecule,
in contrast with the usual rotation diagrams.

With all the above in mind, we have developed a slightly different model
that does not suffer from the above inconveniences. We assume that the molecular
clump is a sphere of radius $\Ro$ with temperature and density given by
the following expressions:
\begin{eqnarray}
T & = & \To \left(\frac{R}{\Ro}\right)^q \label{ett} \\
n & = & \no \left(\frac{R}{\Ro}\right)^p \label{enn}
,\end{eqnarray}
with $R$ being the distance from the centre.
For a given energy level, $J$, the mean column density over the clump
is equal to
\begin{eqnarray}
N_J & = & \frac{4\pi}{\Omc d^2} \int_0^{\Ro} n_j R^2 \d{R} \\
    & = & \frac{4}{\Ro^2} \int_0^{\Ro} \no \left(\frac{R}{\Ro}\right)^p
    g_J \frac{B}{T} \exp\left(-\frac{E_J}{T}\right) R^2 \d{R} \\
    & = & 4 \Ro \no g_J B \int_0^1 \frac{x^p}{T}
           \exp\left(-\frac{E_J}{T}\right) x^2 \d{x}
,\end{eqnarray}
with $x=R/\Ro$, $d$ the distance to the source, and $\Omc=\pi\Ro^2/d^2$ the solid
angle subtended by the clump. The rotational constant, $B$, and $E_J$ are
both expressed in temperature units.  Here we have used Eq.~(\ref{enn})
and the Boltzmann law $n_J=g_J (n/Q) \exp(-E_J/T)$ with the $Q(T)$
partition function. We have also assumed $Q=T/B$ for $T\gg B$,
an approximation that holds for the \CO\ molecule (see Townes \&
Schawlow~\cite{tosh}). However, a similar calculation can be made as long
as $Q\propto T^\gamma$ (in the case of MWH, $\gamma=3/2$).

We treated the two cases $J=0$ and $J>0$ separately. For $J=0$, the ground-state energy is $E_0=0$ and the previous equation simplifies to
\begin{equation}
\frac{N_J}{g_J} = 4 \Ro \no \frac{B}{\To} \int_0^1 x^{2+p-q} \d{x}
                = \frac{4}{(3+p-q)} \Ro \no \frac{B}{\To}.
\end{equation}

For $J>0$, following MWH, we have rewritten the integral
expressing $x$ as a function of $y=T/\To$ by means of Eq.~(\ref{ett}), where
we assume $q<0$:
\begin{eqnarray}
\frac{N_J}{g_J} & = & -\frac{4}{q} \Ro \no \frac{B}{\To} \int_1^\infty
                    y^\frac{3+p-2q}{q} \exp\left(-\frac{E_J}{\To y}\right)\, \d{y} \nonumber \\
                & = &  -\frac{4}{q} \Ro \no \frac{B}{\To}
                       \left(\frac{E_J}{\To}\right)^{-\alpha}
                       \int_0^\frac{E_J}{\To} t^{\alpha-1} \exp(-t)\, \d{t}
,\end{eqnarray}
where we have defined $t=E_J/(\To y)$ and $\alpha=(q-p-3)/q$.

In conclusion, we obtain the following:
\begin{equation}
\frac{N_J}{g_J} = \left\{
\begin{array}{lcl}
\frac{A}{\alpha \Gamma(\alpha)} & \Leftrightarrow & J=0 \\
A \left(\frac{E_J}{\To}\right)^{-\alpha} P\left(\alpha,\frac{E_J}{\To}\right) & \Leftrightarrow & J>0 \\
\end{array}
\right.
\label{enj}
,\end{equation}
where we have defined
\begin{equation}
 A = -\frac{4}{q} \Gamma(\alpha) \, \Ro \, \no \frac{B}{\To}    \label{eaa}
,\end{equation}
and
\begin{equation}
 P(a,x) = \frac{1}{\Gamma(a)} \int_0^x t^{a-1} \exp(-t) \, \d{t}
\end{equation}
is the incomplete gamma function.
It is worth noting that, in the limit $E_J/\To\gg1$, we have recovered the result
of MWH, $N_J/g_J \propto E_J^{-\alpha}$,
because $\lim_{E_J/\To\to+\infty}{P(\alpha,E_J/\To)}=1$.

By fitting Eq.~(\ref{enj}) to the data in a rotation diagram,
one can obtain the parameters $A$, $\To$, and $\alpha$.
Finally, one can relate the mean column density of the molecule, $N$, and
the mass of the clump, $M_{\rm H_2}$, to the values of these three parameters
as follows:
\begin{eqnarray}
 N & = & \frac{1}{\Omc d^2} \int_0^{\Ro} n \, 4\pi R^2\d{R} \nonumber \\
   & = & \frac{4}{p+3}\no\Ro \nonumber \\
   & = & \frac{A}{(\alpha-1)\Gamma(\alpha)}\frac{\To}{B}   \label{enco} \\
 M_{\rm H_2} & = & \mu m_{\rm H} \, \Omc d^2 \frac{N}{X}   \label{emass}
,\end{eqnarray}
where $X$ is the abundance of the molecule relative to H$_2$ and we have
used Eq.~(\ref{eaa}) to express $\no$ as a function of $A$.

\end{appendix}

\end{document}